\documentclass[prl,nobalancelastpage,twocolumn,nofootinbib,superscriptaddress,showpacs]{revtex4-2}

\usepackage{color,amsthm,amsmath,amsfonts,graphicx,bm}
\usepackage{mathtools, nccmath}
\usepackage{bm}
\usepackage{bm,eufrak}
\usepackage[export]{adjustbox}
\usepackage{amsfonts}
\usepackage{amssymb}
\usepackage{amsmath,mathrsfs}
\usepackage{epsfig}
\usepackage{graphicx}
\usepackage{enumerate}
\usepackage{multirow}
\usepackage{verbatim}
\usepackage{subfigure}
\usepackage{bbold}
\usepackage{soul}
\usepackage{scrextend}
\usepackage{tikz}
\usepackage{comment}
\allowdisplaybreaks
\usepackage[linesnumbered,ruled,vlined]{algorithm2e}
\usepackage{algorithmic}

\usepackage{ragged2e}

\newcommand{\mc}[1]{\mathcal{#1}}
\newcommand{\mr}[1]{\mathrm{#1}}
\newcommand{\dg}{\dagger}

\newcommand{\mb}{\mathbf}

\renewcommand{\O}{\mathcal{O}}

\newcounter{notes}
\stepcounter{notes}

\DeclareFontFamily{OT1}{pzc}{}
\DeclareFontShape{OT1}{pzc}{m}{it}{<-> s * [1.10] pzcmi7t}{}
\DeclareMathAlphabet{\mathpzc}{OT1}{pzc}{m}{it}

\newtheorem{theorem}{Theorem}

\newtheoremstyle{named}{}{}{\itshape}{}{\bfseries}{.}{.5em}{\thmnote{#3}}
\theoremstyle{named}
\newtheorem*{namedtheorem}{Theorem}

\begin{document}

%\title{Complexity bounds on tensor network contractions and quantum circuit simulations}
%\title{Simulating quantum circuits using tensor network contractions with analytical guarantees}

\title{Simulating quantum circuits using efficient tensor network contraction algorithms with subexponential upper bound}

\date{\today}

\author{Thorsten B. Wahl}
\affiliation{Department of Applied Mathematics and Theoretical Physics, University of Cambridge, Wilberforce Road, CB3 0WA, United Kingdom.}
\author{Sergii Strelchuk}
\affiliation{Department of Applied Mathematics and Theoretical Physics, University of Cambridge, Wilberforce Road, CB3 0WA, United Kingdom.}

\begin{abstract}
	%{\color {blue} TNs are ubiquitous of the leading toolkits...}
	We derive a rigorous upper bound on the classical computation time of finite-ranged tensor network contractions in $d \geq 2$ dimensions. %By means of the Sphere Separator Theorem, we  are able to take advantage of the structure of quantum circuits to speed up contractions to show that  
	%a tensor network can be contracted in subexponential time in the number of tensors if these have a finite-ranged connectivity. Using these findings, we rigorously demonstrate that 
	Consequently, we show that 
	quantum circuits of single-qubit and finite-ranged two-qubit gates can be classically simulated in subexponential time in the number of gates. Moreover, we present and implement an algorithm guaranteed to meet our bound and which finds contraction orders with vastly lower computational times in practice. In many practically relevant cases this beats standard simulation schemes and, for certain quantum circuits, also a state-of-the-art method. Specifically, our algorithm leads to speedups of several orders of magnitude over naive contraction schemes for two-dimensional quantum circuits on as little as an $8 \times 8$ lattice. We obtain similarly efficient contraction schemes for Google's Sycamore-type quantum circuits, instantaneous quantum polynomial-time circuits, and non-homogeneous (2+1)-dimensional random quantum circuits. 
	
	%We explicitly consider several such examples, for which an algorithm based on the sphere separator theorem leads to speedups of several orders of magnitude over naive contraction schemes already for an $8 \times 8$ lattice. 
\end{abstract}

\maketitle

\textit{Introduction. --} 
Tensor network methods are a single most impactful toolkit responsible for some of the most dramatic improvements in a large number of areas of physics and computer science. They revolutionized our understanding of condensed matter physics~\cite{Eisert2013,Orus2019}, lattice field theories~\cite{Tagliacozzo2014,Silvi2014,Zohar2015,Zohar2016,Banuls2018}, quantum information and computing~\cite{Markov2008,Ferris2014,Guo2019,Pan2020,Huang2021,Farrelly2021,Levental2021,Napp2022}, and machine learning~\cite{Stoudenmire2016}, achieving results which are far out of reach for analytical methods. Recently, Google exhibited a quantum computational device capable of reaching quantum supremacy -- by presenting a highly unstructured computational task which would reportedly take an exceptionally long time on a classical supercomputer to solve~\cite{Arute2019googlesycamore}. In the absence of a formal proof of hardness, this challenge spurred a number of exciting developments on the classical algorithms side with the aim to find an efficient solution by classical means~\cite{Guo2021,Gray2021,Alibaba}. Tensor network contraction methods coupled with ingenious empirical contraction strategies exhibited their immense power reducing the classical computational time to mere hours and even minutes~\cite{pan2021solving,pan2021simulating,liu2021closing}.

The quest for an efficient solution using tensor networks allows one to develop new contraction techniques and intuition about the problem instance. Alas, the resulting simulation algorithm does not typically allow us to solve generic problems in this class efficiently: a slightly tweaked computational task -- whilst still being easy for a quantum computer -- can invalidate the speedup obtained by classical simulation algorithms in this case. This presents one of the major problems when applying tensor network methods: the lack of explicit analytical guarantees on contraction complexity that work well for a range of problems in the class. Efficient empirical approaches that work well for a particular problem instance may comprehensively fail, forcing one to start the search for an efficient simulation method from a clean slate by a costly path of trial and error. Furthermore, non-trivial analytical runtime guarantees which have been found thus far depend on quantities that are themselves hard to compute~\cite{treewidth1,treewidth2}. 

Clearly, little can be done analytically when the underlying problem is unstructured, i.e. the quantum circuit is made up of uniformly random gates acting on a random subset of qubits. However, as noted in Ref.~\cite{ayral2022density}, this model is not representative of practical quantum computational processes -- quantum algorithms expressed in the circuit model yield circuits which are far from uniformly random. Considering circuit topology and the `macroscopic' structure (or layout) of quantum circuits can provide us with valuable extra information which may subsequently be used to derive efficient quantum simulation algorithms. We develop a method which allows us to take advantage of the circuit structure and thus for the first time yields tensor network contraction algorithms with nontrivial theoretical runtime guarantees. This method is based on the idea contained in a rich class of so-called separator theorems~\cite{Lipton1979planar, miller1997separators}. In their simplest form, they represent isoperimetric inequalities for planar graphs. Such (planar) separator theorems state that any planar graph can be split into smaller subgraphs by removing a fraction of its vertices. 
%More precisely, removing $O(\sqrt{n})$ vertices from a graph with $n$ vertices partitions it into disjoint subgraphs each of which has at most $2n/3$ vertices. 
The Planar Separator Theorem~\cite{Lipton1979planar} has found uses in classical complexity theory -- counting satisfiability problems (\#SAT) problems and \#Cubic-Vertex-Cover, where it was decisively better than the state-of-the-art solvers~\cite{Kourtis2019couting} and Boolean symmetric functions~\cite{Gray2021}, and has been mentioned in the context of quantum circuits~\cite{Liu2020}. We apply the ideas outlined in separator theorems to derive analytical upper bounds on the classical simulation times of quantum circuits taking into account the explicit layout of each of the quantum circuits. Recently, the authors of Ref.~\cite{huang2020explicit} argued that (assuming the Strong Exponential Time Hypothesis) strongly simulating certain $poly(n)$ depth quantum circuits on $n$ qubits requires a $2^{n-o(n)}$ time using tensor network methods. Our work rigorously demonstrates how one can derive a constructive tensor-contraction method with subexponential upper bounds on its runtime if we take advantage of extra information about the structure of the circuit. 

%can circumvent the limitations of the tensor-network techniques when one incorporates the structural features of the circuit to derive a constructive tensor-contraction method with subexponential upper bounds and on its runtime.

This Letter is structured as follows: We first give a very brief introduction to the types of tensor network contractions that appear in the range of practical above applications, emphasizing their `metadata' which will be incorporated in our algorithm. Thereafter, we recall the Sphere Separator Theorem~\cite{miller1997separators} and use it to prove a rigorous upper bound on the classical contraction time of $d \geq 2$-dimensional tensor networks. Then, we demonstrate the power of this result %(alongside the discussion of a class of Planar Separator Theorems~\cite{Lipton1979planar}) 
for quantum circuit simulations, obtaining analytical guarantees on their classical simulation times. 
%We review an algorithm~\cite{miller1997separators} (``Sphere Separator Algorithm'', SSA) that can be used to efficiently determine the contraction order underlying the Sphere Separator Theorem in the Supplemental Material (SM). 
Finally, we consider several examples, for which we numerically demonstrate the advantage provided by our method over naive contraction schemes and in certain cases over a state-of-the-art approach (i.e., Cotengra~\cite{gray2020hyper}).

%analytical guarantees over naive contraction schemes for sufficiently large systems that arise in practical applications. 
%We find that our powerful technique based on the Sphere Separator Algorithm (SSA), described in the Supplemental Material (SM), leads to tremendous speedups already for modest system sizes.

\textit{Tensor network contractions. --} 
Below, we provide a short overview of the conventional graphical representation of tensor networks and propose an alternative one, useful for our purposes. Based on the latter, we then employ the Sphere Separator Theorem to prove an upper bound on the classical contraction time of finite-ranged tensor networks in $d \geq 2$ dimensions to a scalar: %, which allows us to calculate relevant expectation values: 
\begin{namedtheorem}[Sphere Separator Theorem (SST)~\cite{miller1997separators}] 
	For a set of $n$ spheres in $d$ dimensions such that each point is contained in at most $k$ spheres the following holds: There exists a sphere $S$ such that removing the set $\Gamma_O(S)$ of spheres which $S$ intersects with gives rise to two mutually non-intersecting sets of spheres $\Gamma_E(S)$ and $\Gamma_I(S)$  with 
	\begin{align}
		|\Gamma_O(S)| &\leq c_{d} k^{1/{d}} n^{1-1/{d}}, \label{eq:O} \\ 
		|\Gamma_E(S)|, |\Gamma_I(S)| &\leq \frac{d+1}{d+2}n. \label{eq:EI}
	\end{align} 
	The coefficients are $c_1 = 1$, $c_2 = 2$, $c_3 < 2.135$, $c_4 < 2.280$, $c_5 < 2.421$, and in general $c_d < \sqrt{2 d/\pi} [1 + \mathcal{O}(1/\log(d))]$ for $d > 1$~\cite{Smith1998cubesep}. 
\end{namedtheorem}
Notably, the proof of the SST is constructive and there exists a randomized algorithm to calculate the above separator in polynomial time, which we call the Sphere Separator Algorithm (SSA). We reproduce its steps from Ref.~\cite{miller1997separators} in the Supplemental Material~(SM). 

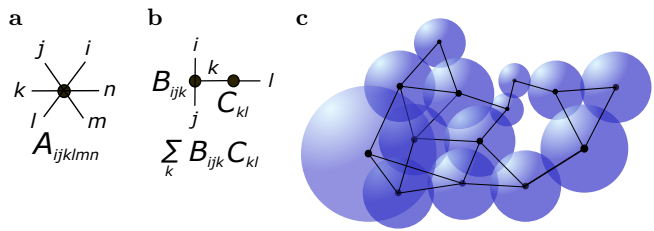
\begin{figure}[t]
	\begin{picture}(40,85)
		\put(0,25){\includegraphics[width=0.075\textwidth]{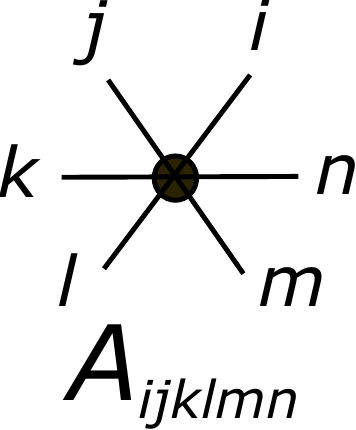}}
\put(-2,77){\textbf{a}}
	\end{picture} \ \ \
\begin{picture}(50,85)
		\put(0,20){\includegraphics[width=0.09\textwidth]{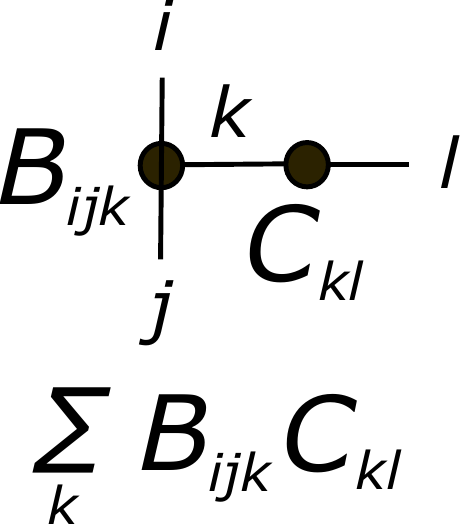}}
\put(-2,77){\textbf{b}} 
\end{picture} \ 
\begin{picture}(130,85)
		\put(0,0){\includegraphics[width=0.26\textwidth]{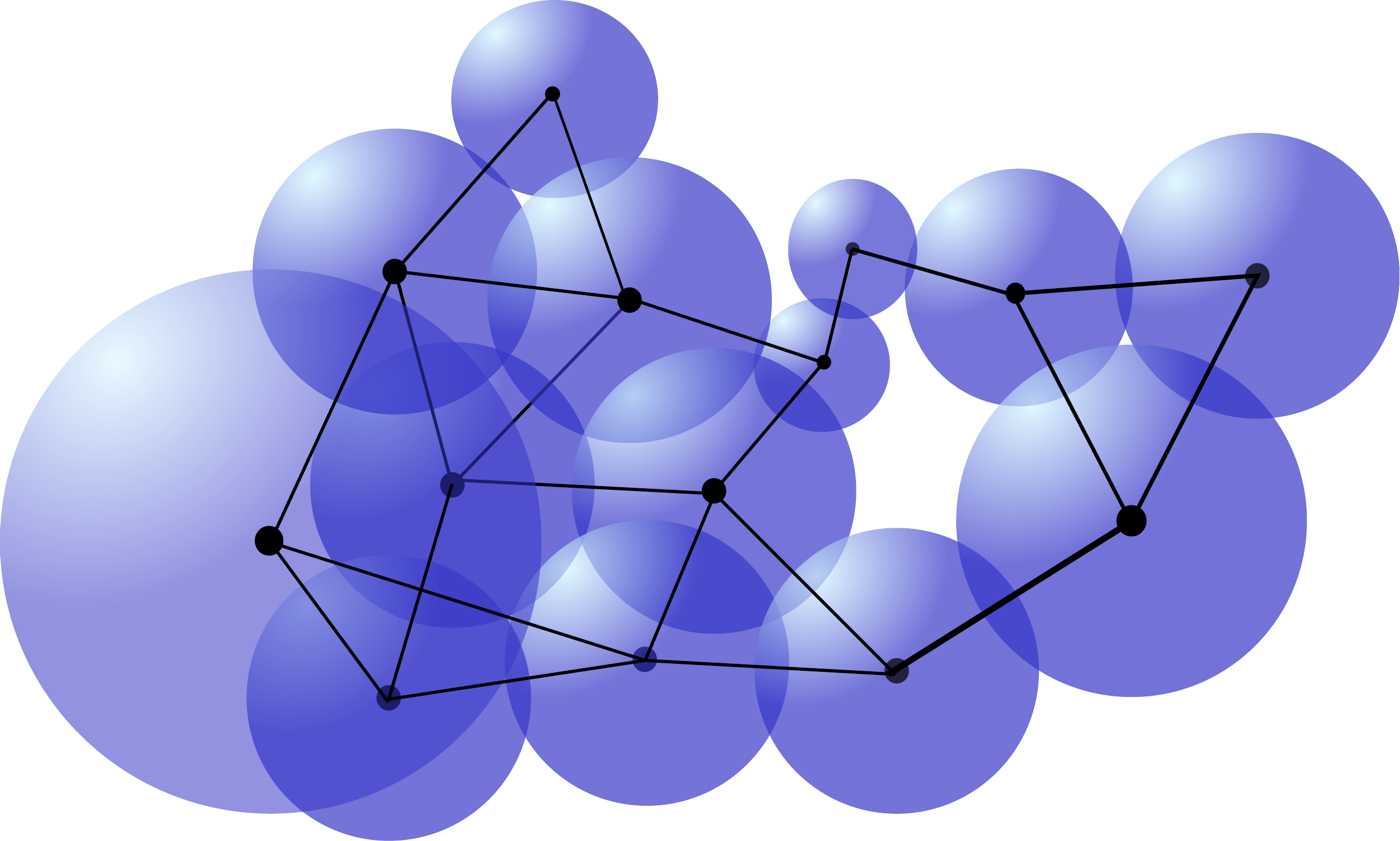}}
\put(-2,77){\textbf{c}}
\end{picture}
	\caption{a: Conventional graphical representation of a tensor and (b) a contraction of two tensors. c: New graphical representation, where each tensor is endowed with a sphere, and only intersecting spheres \textit{can} correspond to tensors with a contracted index. In the shown example, up to $k = 3$ spheres overlap at any point.}    
	\label{fig:spheres}
\end{figure}

In applications of tensor networks, their graphical short-hand notation has become indispensable. Conventionally, a rank-$m$ tensor is represented by a box or a sphere (or, as in our work, by a dot) with $m$ emanating lines. Each line corresponds to one tensor index and a line connecting two tensors to a contraction (i.e., a summation) of the corresponding index, see Fig.~\ref{fig:spheres}a,b. 
%We consider the contraction  of a $d$-dimensional tensor network ($d \geq 2$) of $n$ tensors to a scalar
%with the following theorem as our main tool:  
In order to take advantage of the SST, we convert the conventional graphical representation using the following steps: (i)~Find an embedding of the tensor network graph into $\mathbb{R}^d$ such that connected tensors are close. (ii)~Represent each tensor by a sphere of radius large enough such that if two tensors are connected, their spheres intersect (though intersecting spheres do not have to correspond to a bond), see Fig.~\ref{fig:spheres}c. Crucially, by choosing the radii of the spheres large enough, one can always ensure that all connected tensors correspond to intersecting spheres. 
\begin{theorem}\label{thm:general}
We consider the full contraction of a tensor network embedded into $\mathbb{R}^d$ ($d \geq 2$) of $n$ tensors of at most $M$ entries each. We assume that there exists a graphical representation of the tensors as spheres of radii such that the spheres of tensors with a contracted index intersect and the maximum number $k$ of spheres overlapping in any point is $n$-independent (finite-ranged tensor network). Then, for sufficiently large $n$, the number of scalar operations to (classically) contract the tensor network is upper bounded by $2n^{1/\log_2(\frac{d+2}{d+1})} M^{a_d k^{1/d} n^{1-1/d}}$, where $a_d = c_d /\left[2-2\left(\frac{d+1}{d+2}\right)^{1-1/d}\right]$.
\end{theorem}

\begin{figure}
	\includegraphics[width=0.33\textwidth]{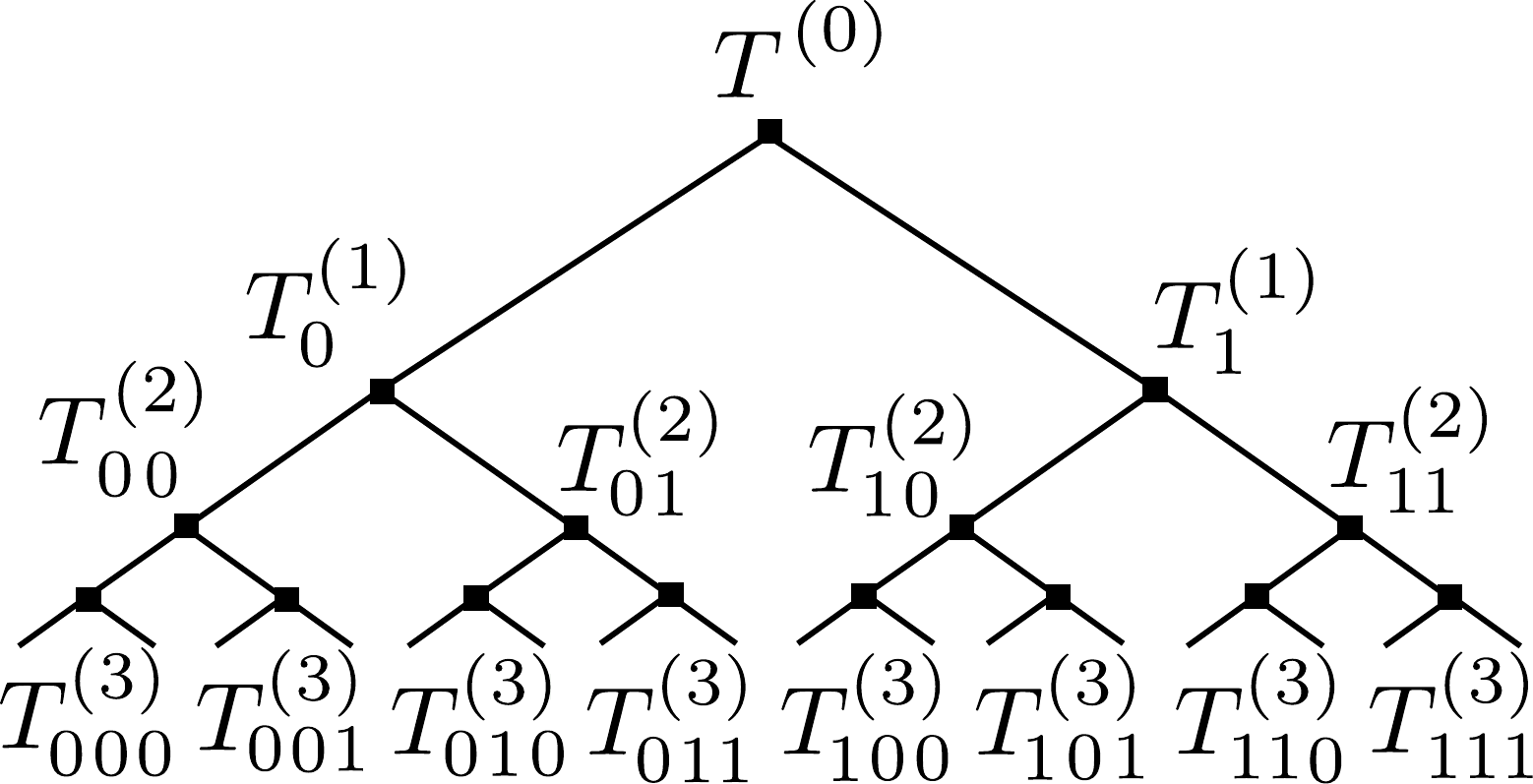}
	\caption{Separator hierarchy: The original tensor network $T^{(0)}$ gets successively split into smaller tensor networks $T^{(\ell)}_{i_1 \ldots i_\ell}$.}    
	\label{fig:hierarchy}
\end{figure}

%This will generally increase the number $k$; however, the assumptions of Theorem~\ref{thm:general} imply that the radii can be chosen to be $\mathcal{O}(1)$ (not growing with the system size $n$), as $k$ is $n$-independent. \\
\textbf{Proof:} We use the SST to split the tensor network into two disconnected tensor networks of at most $n(d+1)/(d+2)$ tensors each [corresponding to $\Gamma_{E,I}(S)$] and the tensors sitting at the interface $\Gamma_O(S)$. Each of these tensors is then included into the one of the earlier two tensor networks with whom it shares indices of higher overall bond dimension. (The choice is arbitrary if the overall connecting bond dimensions are equal for both.) This ensures that each of the assigned tensors shares only indices of overall bond dimension $\leq M^{1/2}$ with the tensor network it has \textit{not} been assigned to. Thus, the resulting two parts are connected by bonds of overall dimension $\leq~M^{|\Gamma_O(S)|/2}$.   
We recursively apply this procedure to split the tensor networks further until all tensor networks are of size $\O(1)$. 
Let $T_{i_1 i_2 \ldots i_\ell}^{(\ell)}$ be the corresponding tensor networks, where $\ell$ is the level of the resulting separator hierarchy and the indices $i_j = 0,1$ indicate the path taken through the separator hierarchy, see Fig.~\ref{fig:hierarchy}. The number of scalar operations (multiplications, additions) needed to create $T_{i_1 i_2 \ldots i_\ell}^{(\ell)}$ is bounded by
\begin{align}
t_{i_1 \ldots i_\ell}^{(\ell)} \leq t_{i_1 \ldots i_\ell 0}^{(\ell+1)} + t_{i_1 \ldots i_\ell 1}^{(\ell+1)} + 2 M^{(\ell+1)}_{i_1 \ldots i_\ell},
\label{eq:t_recursive}
\end{align}
where $M^{(\ell+1)}_{i_1 \ldots i_\ell}$ denotes the product of the dimensions of all of the indices of the tensors $T_{i_1 \ldots i_\ell0}^{(\ell+1)}$ and $T_{i_1 \ldots i_\ell1}^{(\ell+1)}$, counting shared indices once. This product equals the total number of scalar multiplications required to contract the two tensors. Since the corresponding products have to be added up, the number of required additions is upper bounded by $M^{(\ell+1)}_{i_1 \ldots i_\ell}$, hence the prefactor of 2 in Eq.~\eqref{eq:t_recursive}. By the SST, $M^{(\ell+1)}_{i_1 \ldots i_\ell } \leq M^{\frac{1}{2}\sum_{j=1}^{\ell+1} c_d k^{1/d} \left [n(\frac{d+1}{d+2})^{j-1}\right]^{1-1/d}}$, as the sum in the exponent is the maximum number of constituting tensors which can sit at the combined surface %(including the separating cut) 
of the tensor networks $T_{i_1 \ldots i_\ell0}^{(\ell+1)}$ and $T_{i_1 \ldots i_\ell1}^{(\ell+1)}$ (counting the surface separating them once), and $M^{1/2}$ is the maximum bond dimension per tensor. Repeatedly inserting Eq.~\eqref{eq:t_recursive} into itself yields the closed form
\begin{align}
t^{(0)} &\leq  \sum_{\ell = 0}^{z-1} 2^{\ell+1} M^{\frac{1}{2}\sum_{j=1}^{\ell+1} c_d k^{\frac{1}{d}} [n(\frac{d+1}{d+2})^{j-1}]^{1-\frac{1}{d}}} + 2^{z} M^{\O(1)}, \label{eq:t0_closed}
\end{align}
where $z \leq 1+\log_2(n) / \log_2 (\frac{d+2}{d+1})$. %is the level of the separator hierarchy where generating the constituting tensor networks from scratch by brute-force contraction will be computationally cheaper [leading to the right term in Eq.~\eqref{eq:t0_closed}] than the cost given by the bound in Eq.~\eqref{eq:O}. 
After upper bounding the sum in the exponent by a geometric sum, we obtain
\begin{align}
t^{(0)} < 2 n^{1/\log_2 (\frac{d+2}{d+1})} M^{\frac{c_d k^{1/d}}{2- 2(\frac{d+1}{d+2})^{1-1/d}} n^{1-1/d}} \label{eq:t0_final}
\end{align}
for sufficiently large $n$. 
%By the definition of $z$, we have $c_d k^{1/d} [n (\frac{d+1}{d+2})^z ]^{1-1/d} = n (\frac{d+1}{d+2})^{z+1}$, and thus $z = \log_2(\frac{n}{c_d^d k})/\log_2(\frac{d+2}{d+1}) - d$, giving the stated result. 
\qed

When the underlying connectivity graph is planar, a more appropriate tool is an analogue of the SST -- the Planar Separator Theorem~\cite{Lipton1979planar}:
\begin{namedtheorem}[Planar Separator Theorem (PST)~\cite{Lipton1979planar}]
A planar graph, i.e., a two-dimensional graph of non-intersecting lines and $n$ vertices, can be separated into two disconnected graphs of at most $2n/3$ vertices each by removing $c_{PST} \sqrt{n} + \mathcal{O}(1)$ vertices with $c_\mr{PST} < 1.971$~\cite{Djidjev1997improved}. 
\end{namedtheorem}
We note that in Refs.~\cite{Kourtis2019couting,Gray2021} the overall exponent of $M$ was not calculated; in particular, it was not clear if it is still of order $\O(\sqrt{n})$ when one takes the entire separator hierarchy into account. After evaluating the expression corresponding to Eq.~\eqref{eq:t0_closed}, we obtain an improved upper bound of $t^{(0)} < 2n^{1/\log_2(3/2)}M^{a_2' \sqrt{n}}$ with $a_2' = c_\mr{PST} /(2-2\sqrt{2/3})  < a_2 = c_2/(2-2\sqrt{3/4})$ for $d = 2$.

\begin{figure}
%	\begin{picture}(150,50)
%		\put(20,5){\includegraphics[width=0.2\textwidth]{tensors.pdf}}
%				\put(-15,35){\textbf{a}}
%	\end{picture} \\
\begin{picture}(100,120)
		\put(0,0){\includegraphics[width=0.22\textwidth]{collapse.pdf}}
\put(0,112){\textbf{a}} 
\end{picture} \ \ \ \ \ \ \ \ \
\begin{picture}(100,120)
		\put(0,2){\includegraphics[width=0.2\textwidth]{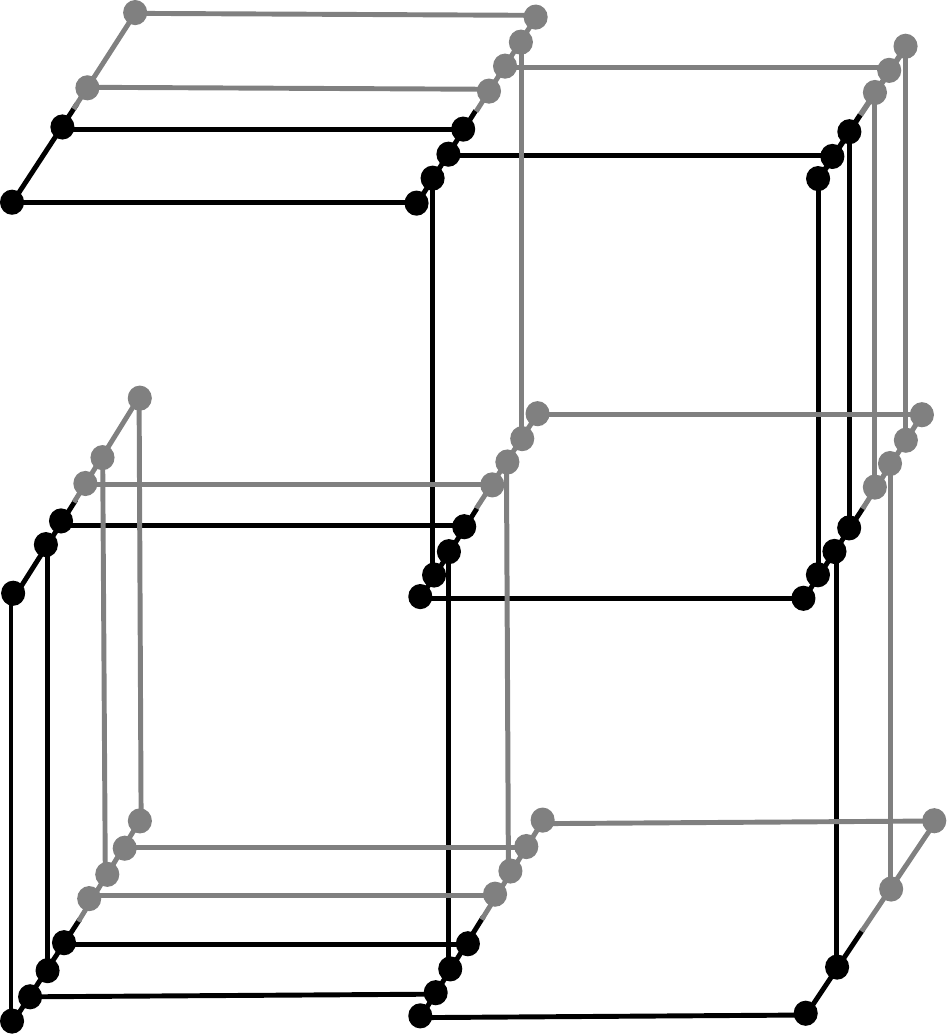}}
\put(-3,112){\textbf{b}}
\end{picture}
	\caption{%a: Singular value decomposition of the tensor corresponding to a two-qubit gate into two rank-3 tensors (represented as dots), where the diagonal matrix containing the singular values has been absorbed into one of the rank-3 tensors. The connecting line has  dimension 4, while all other lines represent dimension 2 indices. 
	a: Collapse of the $(d+1)$-dimensional tensor network corresponding to a quantum circuit onto $d$-dimensional space. In the Figure, $d = 2$, and the qubits are arranged on a $3 \times 3$ square lattice (dashed lines) with nearest-neighbor gates. % The tensor network contraction is graphically represented both in the conventional way and in the way introduced in this work. 
Note that the collapse of the time dimension causes some tensors (and spheres) to lie on top of each other. 
	b: Collapse of the $(d+1)$-dimensional tensor network onto $d$-dimensional space with the time direction corresponding to slightly offset tensors. %, shown in the conventional graphical representation. 
	Tensors coming from the (adjoint) unitary $U^{(\dg)}$ are shown in black (gray).} %In order to convert this graphical representation to a planar graph, vertices have to be placed at the intersection points.}    
	\label{fig:tensors}
\end{figure}

%\textit{Quantum circuit simulations. --}
\textit{Classical Simulation of Quantum Circuits. --} We now apply the SST and  PST to derive analytical guarantees on the classical simulation time of quantum circuits, taking into account the explicit layout of each of them. 
Consider a quantum circuit $U$ of single-qubit and finite-ranged two-qubit gates acting on $N$ qubits over $T$ time steps. For simplicity, we assume that the system is initialized in the $|0 0 \ldots 0 \rangle := |\mb{0}\rangle$ state. We represent the expectation value $c$ of the measurement $P = \bigotimes_i P_i$, where $P_i$ is a projector on qubit $i$, as a tensor network, $c = \langle \mb{0} | U^\dg \bigotimes_i P_i U | \mb{0} \rangle$. 
 Two-qubit gates correspond to rank-4 tensors and single-qubit gates to rank-2 tensors (unitary matrices). The latter as well as the $P_i$ can be absorbed into the former, such that only the two-qubit gates affect the scaling of computational complexity: 
\begin{theorem}\label{thm:flat} 
The number of scalar operations to classically simulate a $(d+1)$-dimensional quantum circuit with $d\ $$\geq\ $$2$  of single- and two-qubit gates of maximum range $l$, acting on a set of $N$ qubits of minimal distance $r$ apart, is upper bounded by $\left[2\sum_{i=1}^N f(\mb x_i)\right]^{1/\log_2 (\frac{d+2}{d+1})} 2^{1+8 a_d (1+l/r)F^{1/d} \left[\sum_{i=1}^N f(\mb x_i)\right]^{1-1/d}}$ for sufficiently large $N$.  $f(\mb x_i)$ denotes the number of two-qubit gates acting on the qubit $i$ at position $\mb x_i$\ $\in$\ $\mathbb{R}^d$ and $F = \max_i f(\mb x_i)$. 
%For $d = 1$ the upper bound is $N \, 2^{4F}$, where $N$ is the number of qubits. 
For $d = 2$ and nearest neighbor two-qubit gates ($l = r$), a tighter bound is $[\sum_{i=1}^N f(\mb x_i)(2+  \frac{1}{2} f(\mb x_i))]^{1/\log_2 (3/2)} 2^{1+4a_2' \sqrt{\sum_{i=1}^N f(\mb x_i) [2+f(\mb x_i)/2]}}$.
\end{theorem}
\textbf{Proof:} We collapse the time dimension to length zero, such that the positions of all tensors are given by spatial coordinates only. We split the two-qubit gates using a singular value decomposition into a contraction of two rank-3 tensors connected by a bond of dimension 4. 
Hence, each gate acting on qubits $i$ and $j$ gives rise to two tensors, which we place at positions $\mb x_i + (\mb x_j - \mb x_i)/4$ and $\mb x_j + (\mb x_i - \mb x_j)/4$, respectively. We choose the spheres of the tensors to have radius $l/4 + \epsilon$ ($\epsilon > 0$), such that all connected tensors correspond to intersecting spheres, see Fig.~\ref{fig:tensors}a.  
In Theorem~\ref{thm:general}, we have $n = 2 \sum_{i=1}^N f(\mb x_i)$ tensors (coming from $U$ and $U^\dg$). Each point in space is only covered by spheres whose origins are at a distance $\leq l/4 + \epsilon$ (their radius). As $r/2$ is the minimal distance between the centers of the spheres coming from different qubits (we moved the tensors 50\% closer together than the qubits), we therefore have $k < 2 F (\frac{l/4+r/4}{r/4})^d$, where the third factor upper bounds the number of solid balls of radius $r/4$ contained in a sphere of radius $l/4 + r/4$. With Theorem~\ref{thm:general} and since the number of entries of each tensor is $M = 16$, we obtain the stated scaling.  
For $d = 2$ and nearest-neighbor two-qubit gates, we can use the PST after transforming our graph to one whose vertices are connected by non-intersecting lines: Going  back to the conventional graphical representation, we displace the tensors around $\mb x_i$ corresponding to the same qubit $i$ but to different times by $\epsilon' > 0$ with respect to each other, see Fig.~\ref{fig:tensors}b. In the limit $\epsilon' \rightarrow 0$ there can be up to $[f(\mb x_i)]^2/2$ intersections between the lines connecting the tensors around the qubit at $\mb x_i$ to tensors around neighboring qubits [the lines of $f(\mb x_i)/2$ gates applied at an earlier time intersecting with $f(\mb x_i)$ lines of gates from a later time and the corresponding adjoint operation]. We can transform this graph to a planar graph by placing vertices (identity tensors) at the corresponding intersection points, resulting in $n \leq \sum_{i=1}^N \left(2f(\mb x_i) + [f(\mb x_i)]^2/2\right)$ vertices overall. The identity tensors placed at the intersection points have $M = 4^4$ entries, and we thus obtain the stated tighter bound. 
\qed

As $\sum_{i=1}^N f(\mb x_i) \leq N F$, the above bounds generally beat left-to-right contraction: for example, for a hypercubic $L^d$ lattice with nearest neighbor two-qubit gates, left-to-right contraction yields a bound $t_\mr{side} \leq 2^{1+4FL^{d-1}} L = 2^{1+4FN^{1-1/d}} N^{1/d}$. This is larger than the bounds of Theorem~\ref{thm:flat} for a very inhomogeneous $f(\mb x)$, where the gates are applied in a very irregular fashion (not to be confused with the set of gates that could in principle be applied, which is often regular~\cite{Arute2019googlesycamore}). This is, for instance, commonly the case in quantum error correction applications, where errors to be corrected occur in a spatially irregular fashion. Similarly, the above bounds improve over the classical simulation time of explicit time evolution $t_\mr{expl} \leq T N 2^{2+N}$ if $F < N^{1/d}/[8a_d (1+l/r)]$. 

\textit{Examples. --} We now consider several quantum circuits acting on $L \times L$ qubits, for which we showcase the strengths of our approach. %and numerically demonstrate the quantitative advantage of using the SSA. 
Specifically, we demonstrate that for short-range instantaneous quantum polynomial-time (IQP) quantum circuits~\cite{Bremner2017} the SSA repeatedly applied to sets of spheres centered at the gate positions (with a collapsed time dimension) gives rise to massively faster contraction orders than suggested by Theorem 2. As a result, the SSA approach outperforms naive contraction schemes and a state-of-the-art method already for small system sizes. Afterwards, we consider Google Sycamore-type~\cite{Arute2019googlesycamore} quantum circuits, where the bound of Theorem 2 outperforms naive contraction schemes for large system sizes and the SSA again already for small system sizes. %, where again the SSA produces dramatically lower computational costs in practice than the bound of Theorem~\ref{thm:flat}. 
Finally, we find that the bound of Theorem~\ref{thm:flat} improves over naive contraction schemes for significantly smaller system sizes for (2+1)-dimensional random quantum circuits with Poisson-distributed cavities. 
%The scaling behavior of the compared methods varies substantially between the considered examples due to the varying inhomogeneity of the  quantum circuits. 

\begin{figure}[t]
\includegraphics[width=0.49\textwidth]{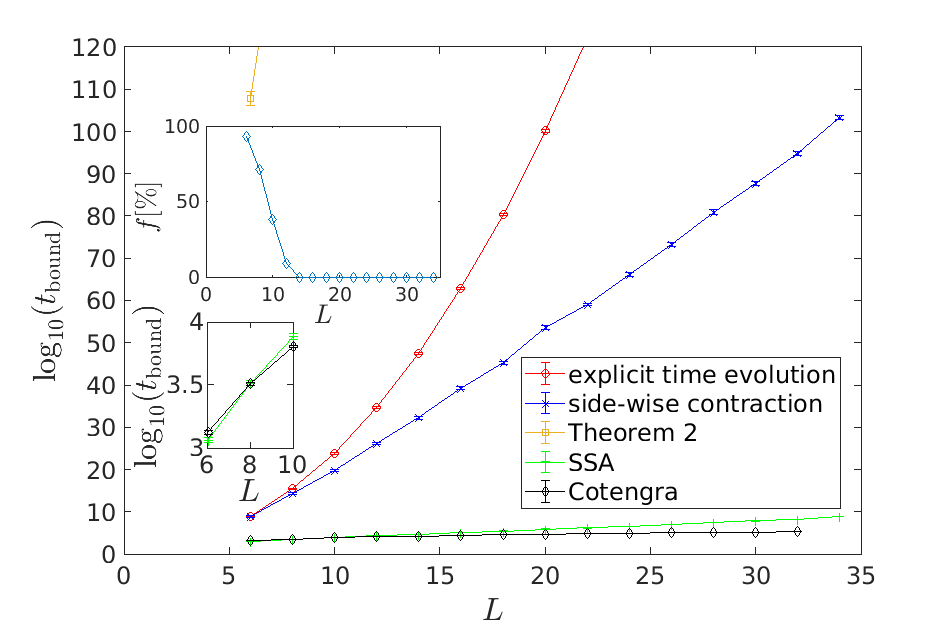}  % figure needs to be updated!
	\caption{Mean logarithms of the bounds obtained for the classical computation time based on explicit time evolution ($T L^2 2^{2+L^2 - \#(\mathrm{idle \ qubits})}$), side-wise contraction ($2^{1+4FL} L$), the second bound of Theorem~\ref{thm:flat}, the bound from the SSA, and the number of scalar operations required by Cotengra~\cite{gray2020hyper}, averaged over 100 quantum circuit realizations of IQP quantum circuits. Top inset: Percentage of realizations for which the SSA (without any additional optimizations) outperforms Cotengra. Bottom inset: Zoom-in of the main figure for small system sizes. Error bars denote the standard error of the mean of the logarithms.}  
	\label{fig:IQP}
\end{figure}

\textbf{1. IQP quantum circuits:} 
Here, we apply the procedure to IQP quantum circuits~\cite{Bremner2017} with single- and two-qubit gates: In each of the $T$ time steps, a single-qubit gate is acted on a qubit with probability $7/8$ [corresponding to a randomly chosen phase gate $\mathrm{diag}(1,e^{i\pi m /4}), \ m \in \{0, 1, \ldots, 7\}$], and afterwards two-qubit gates act on all nearest-neighbor pairs of qubits which have not yet been acted on in this time step with probability $p = 3/4 \cdot \gamma \ln(N)/N$. The prefactor of $3/4$ corresponds to randomly chosen gates $\mathrm{diag}(1,1,1,i^m), \ m \in \{0, 1, 2, 3\}$. The results for $T = L^2$ and $\gamma = 3$ are shown in Fig.~\ref{fig:IQP}. While the bound from Theorem 2 is large, the one from the SSA performs many orders of magnitude better, also beating naive contraction schemes and the state-of-the-art method Cotengra~\cite{gray2020hyper} for most (some) quantum circuit realizations for $L \leq 8$ ($L \leq 12$).

\textbf{2. Google Sycamore-type quantum circuits:} We consider a quantum circuit of the type of Ref.~\cite{Arute2019googlesycamore}, where each qubit (apart from the edge ones) is acted on by one single-qubit gate and one two-qubit gate coupling it to a nearest neighbor per ``cycle''. There are eight cycles of such couplings, which are repeated periodically (i.e., with two two-qubit gates acting on the same nearest neighbor). We assume that there are $q$ such periods of eight cycles and that in each period a given qubit is inaccessible with probability $p$, i.e., no single- or two-qubit gate acts on it for the entire period. 
%This induces enough inhomogeneity such that the second bound of Theorem~\ref{thm:flat} improves over explicit time evolution and the side-wise contraction bound $t_\mr{side} \leq 2^{4FL} L$ for sufficiently large $L$. 
We calculated the corresponding bounds for $q = 5$ and $p = 0.88$ (which is below the percolation threshold~\cite{Kesten1980}) and averaged over 100 quantum circuit realizations for each $L$. The results, presented in detail in the SM, indicate that the bound of Theorem~\ref{thm:flat} outperforms naive contraction schemes for most quantum circuit realizations for $L \geq 1600$. (We found it to be superior for some quantum circuit realizations already for $L \geq 400$.) %Again, the SSA beats the naive contraction schemes already for $L \geq 8$. %, much smaller system sizes than the general bounds of Theorem~\ref{thm:flat} suggest! 
%Hence, while the bound of Theorem~\ref{thm:flat} improves over standard schemes only for large system sizes, the SSA gives an advantage for much smaller (practically relevant) system sizes. Note that while one could adapt the side-wise contraction scheme to the specific cavities in a quantum circuit, the scaling of the bound of the SSA is so much stronger that this is unlikely to have a substantial effect in comparison. 

\textbf{3. (2+1)-dimensional random quantum circuits:} We consider a quantum circuit of $T = \alpha L$ time steps. At each time step, nearest-neighbor gates are densely placed with a random orientation unless there is a cavity in the quantum circuit. These cavities have size $S \times S \times S$ in space-time. Their maximum number $v$ is chosen randomly according to a Poisson distribution with parameter $\lambda$ corresponding to a probability $p_v(\lambda)$. The up to $v$ cavities are placed randomly in the quantum circuit [coordinates $(x,y,t)$]; each one appears with probability $p(x,y,t) = \exp\left[-\left(x^2/L^2 + y^2/L^2 + t^2/T^2\right)/\sigma^2\right]$. The results for $S = 5$, $\alpha = 0.1$, $\sigma = 10$, and $\lambda = 5 \cdot 10^4 (L/200)^{3.5}$ %and generate 100 quantum circuits randomly per system size $L$. The results 
(displayed in the SM) are similar to the ones of the previous example, but with a sharper transition. They %shown in Fig.~\ref{fig:Poisson} 
indicate that the bound of Theorem~\ref{thm:flat} improves over the one for side-wise contraction and explicit time evolution for most quantum circuit realizations already for $L \geq 250$. 
%and for a significant fraction already at $L = 200$. 

\textit{Conclusion.} -- We have used the SST to derive analytical upper bounds on the classical contraction runtime of finite-ranged higher-dimensional tensor networks. %Based on that and the similar Planar Separator Theorem, 
We proved similar upper bounds on the classical simulation time of arbitrary higher-dimensional quantum circuits with single- and finite-ranged two-qubit gates. 
%We also indicated how the SSA can be implemented to explicitly determine the contraction order underlying our complexity bound for quantum circuits. 
While these bounds improve over naive contraction schemes only for relatively large system sizes, we showed that, in practice, far better upper bounds are obtained with the SSA, which also outperform a state-of-the-art method for certain quantum circuits of relevant sizes. %, which also yields an estimate for the contraction complexity of a given quantum circuit. 
Our approach will find important applications in the context of classical benchmarks for quantum simulations and help to determine the regime where they do not meet the criterion of quantum supremacy. We envision particularly powerful classical tensor network contraction schemes as a result of combining our algorithm with other heuristic methods, such as the stem optimization technique of Ref.~\cite{Alibaba}, or the index slicing approach of Ref.~\cite{Huang2021}.

\acknowledgments

TBW was supported by the ERC Starting Grant No. 678795 TopInSy and the Royal Society Research Fellows Enhanced Research Expenses 2021 RF{\textbackslash}ERE{\textbackslash}210299. SS acknowledges support from the Royal Society University Research Fellowship.

%\appendix
%\section{Cube Separator Theorem}
%
%Some mileage might be gained by using the Cube Separator Theorem~\cite{Smith1998cubesep} (CST) in $d = 1$ and $d = 2$, which works similarly, but comes with a guarantee to split the set of isooriented cubes with at most a ratio of 1:2, independently of $d$. For the CST, the corresponding coefficient is 
%\begin{align}
%	b_{d'} = \left(\frac{[1 + 2^{d'}(H_{2d'} - H_{d'} - \frac{1}{2})](2d)!}{d!}\right)^{1/d},	
%\end{align}
%where $H_m = \sum_{j=1}^m j^{-1}$ is the $m$-th harmonic number. 
%For $d = 1,2$, we have $k = 2$ (see Fig.~\ref{fig:cube_overlaps}). The above geometric series becomes $1/(1-2/3) = 3$, such that we obtain a prefactor $a_d' = 3 b_{d+1} 2^{\frac{d+2}{d+1}}$, i.e., $a_1' = 18$, $a_2' = 36.46$. 

\appendix

\begin{center} \textbf{Supplemental Material}
\end{center}

\begin{algorithm}[b]\label{algo}
\caption{\textbf{Sphere Separator Algorithm}~\cite{miller1997separators}}
\begin{algorithmic}[1]
\justifying
\REQUIRE Positions $\mathcal P = \{\mb p_1, \mb p_2, \ldots, \mb p_s\} \in \mathbb{R}^{d}$ of centers of spheres, representing $s$ tensors, and the corresponding radii $\mc{R} = \{r_1, r_2, \ldots, r_s\}$. \\
\STATE Compute $\Pi(\mathcal P) = \{\Pi(\mb p_1), \Pi(\mb p_2), \ldots, \Pi(\mb p_s)\}$, where $\Pi$ denotes the stereographic projection from $\mathbb{R}^{d}$ to the unit sphere $S^{d} \subset \mathbb{R}^{d+1}$. \\
\STATE Compute a centerpoint~\cite{Edelsbrunner1987} $\mb c \in \mathbb{R}^{d+1}$ of $\Pi(\mathcal P)$, i.e., all hyperplanes containing $\mb c$ divide the set of points $\{\Pi(\mb p_1), \Pi(\mb p_2), \ldots, \Pi(\mb p_s)\}$ in a ratio $(d+1) \,$:$\, 1$ or less. $\mb c$ is calculated efficiently by randomly selecting subsets $\mathcal S \subset \Pi(\mathcal P)$ of size $|\mathcal S| = a$ and calculating their centerpoints $\mb c_{\mathcal S}$ using Linear Programming on $\mc{O}(a^{d+1})$ linear inequalities of $d+1$ variables. (For any $\delta > 0$ this approach produces a $(d+1+\delta) \,$:$\, 1$ centerpoint with high probability if $a > g(\delta, d+1)$, where $g$ is an $s$-independent function~\cite{Vapnik1971,Haussler1987,Teng1991}, making the approach scalable.) \\
\STATE Compute an orthogonal matrix $R \in \mathbb{R}^{(d+1) \times (d+1)}$ such that $R \, \mb c = (0,0, \ldots, 0, ||\mb c||)$.  \\
\STATE Define the dilatation map $D_\alpha = \Pi \circ (\alpha \mathbb{1}) \circ \Pi^{-1}$, where $\alpha = \sqrt{(1- ||\mb c||)/(1+ ||\mb c||)}$. \\
\STATE Choose a random great circle $C$ on $S^{d}$. The center of $S^{d}$ can be shown to be a centerpoint of $D_\alpha \circ R \circ \Pi(\mathcal P)$~\cite{miller1997separators}, i.e., $C$ divides these points in a ratio $1\,$:$\, (d+1+\delta)$ or better. $C$ gives rise to a sphere $S \subset \mathbb{R}^{d}$ after transforming back to the original $\mathbb{R}^d$ space, which (generically) satisfies Eq.~(2). \\
\STATE Calculate the sphere $S = \Pi^{-1} \circ R^\top \circ D_{\alpha}^{-1}(C)$. It can be proven that $S$ also satisfies Eq.~(1) with probability at least $1/2$~\cite{miller1997separators} for sufficiently small $\delta > 0$. \\
\STATE Check, taking account of the radii $\mc{R}$, if Eqs.~(1) and~(2) are satisfied. If not, choose another great circle $C' \in S^{d}$ and repeat steps 5 and 6 until successful. (In general, if any of the above approaches is successful with probability  $\rho = \mathcal{O}(1)$, it has to be carried out $1/\rho$ times on average.) \\
\ENSURE Sphere $S$, $\Gamma_O(S)$, $\Gamma_E(S)$, $\Gamma_I(S)$. 
\end{algorithmic}
\end{algorithm}

\section{Numerical implementation}

The proof of the Sphere Separator Theorem (SST), used in the main text, is constructive~\cite{miller1997separators}, and there also exists an algorithm to efficiently calculate the corresponding sphere separator. Here we show how this algorithm can be used to compute a separator hierarchy satisfying the bounds of Theorems~1 and~2. As illustrated with examples in the main text (see also next section), the separator hierarchies obtained in practice correspond to much lower classical runtimes. 

The algorithm to calculate the sphere separators of Theorems~1 and~2 is a probabilistic approach and was first presented in Ref.~\cite{miller1997separators}. In our case,  Algorithm~\ref{algo} (summarized above) has to be applied  $\mathcal{O}(n)$ times, where $n$ is the number of tensors. 
%to generate the separator hierarchy underlying the simulation times stated in Theorems~\ref{thm:general} and~\ref{thm:flat}.
The separator hierarchy (see main text) allows us to provide an upper bound on the classical simulation time of the specific tensor network Algorithm~\ref{algo} is applied to: At the lowest level of the separator hierarchy are clusters of few tensors.  The contraction of each of these clusters contributes $\mathcal{O}(1)$ to the overall contraction time. For a given tensor network and computed separator hierarchy, we use a greedy algorithm to upper bound the contraction time of the small clusters to individual tensors at the bottom of the separator hierarchy. At all higher levels, the separator hierarchy uniquely specifies in which order the corresponding tensors have to be pairwise contracted, such that the corresponding computational times can be calculated explicitly. %(assuming the tensors are not sparse). 
The obtained upper bound for the overall contraction time is at least as good as the general bound of Theorem~2, but, in general, many orders of magnitude better.

%If we go down the separator hierarchy and denote by $S_j$ the sphere with the largest $|\Gamma_O(S)|$ at the $j$-th level of the separator hierarchy, then in analogy to Eq.~(4) in the main text, the computational time can be bounded according to 
%\begin{align}
%t_0 \leq  \sum_{\ell=0}^{z-1} 2^{\ell} M^{\frac{1}{2}\sum_{j = 1}^{\ell+1} |\Gamma_O(S_j)|} + \sum_{m=1}^{2^z} M^{n_m/2}, \label{eq:heuristic}
%\end{align}
%where $n_m$ denotes the number of constituting tensors contained in the $m$-th tensor network at the $z$-th level of the separator hierarchy. For the bounds in the main text we used a refined version of Eq.~\eqref{eq:heuristic}, where we explicitly calculated the bond dimensions separating two regions found by the algorithm.  
%In practice, one could also use Eq.~(3) in the main text directly to obtain an even better upper bound for $t_0$. The specific bound~\eqref{eq:heuristic} for Algorithm~\ref{algo} is at least as good as the bound of Theorem~2, but in most cases many orders of magnitude better. 

\section{Additional numerical data}

\begin{figure}
\begin{picture}(225,175)
		\put(-10,2){\includegraphics[width=0.52\textwidth]{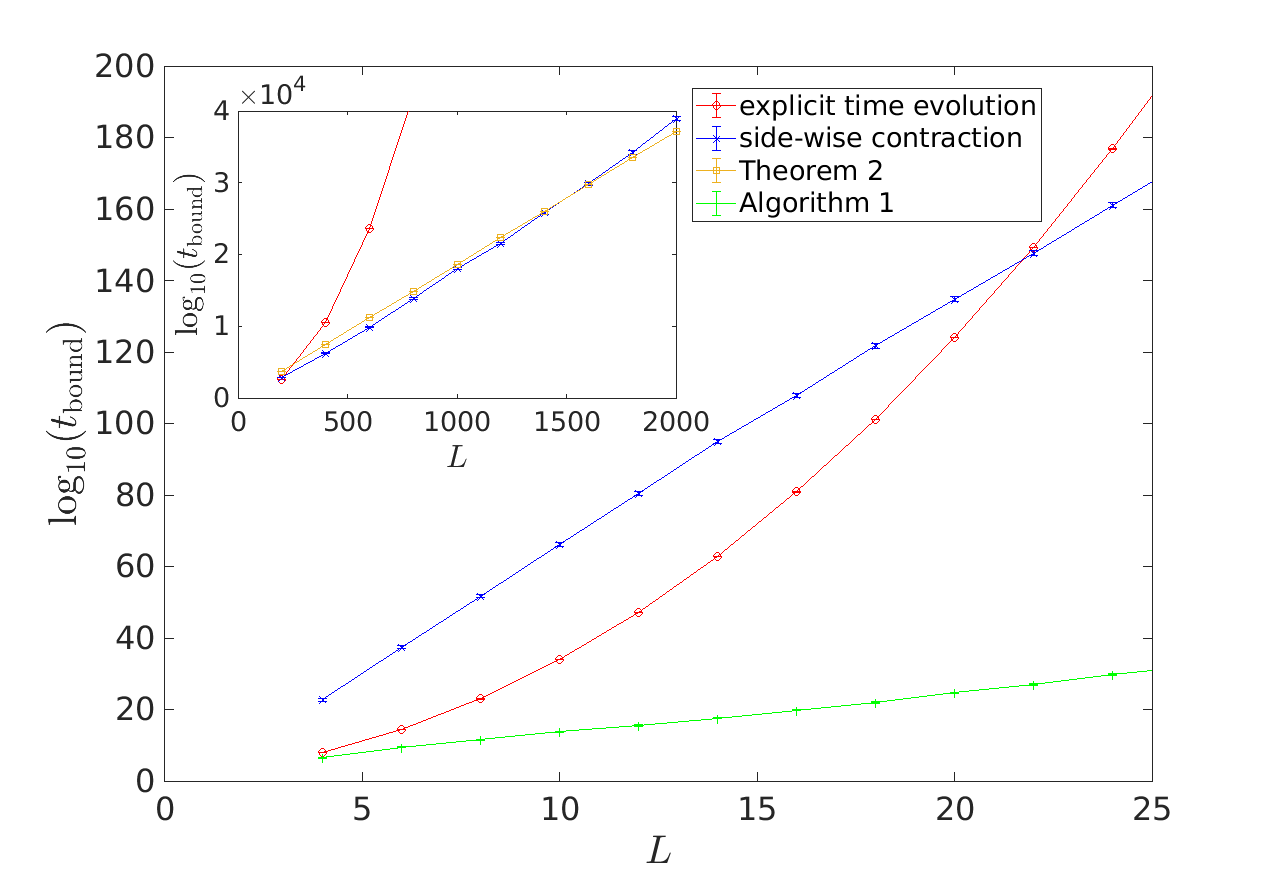}}
\put(-7,175){\textbf{a}} 
\end{picture} \\
\vspace{10pt}
\begin{picture}(225,175)
		\put(-10,2){\includegraphics[width=0.52\textwidth]{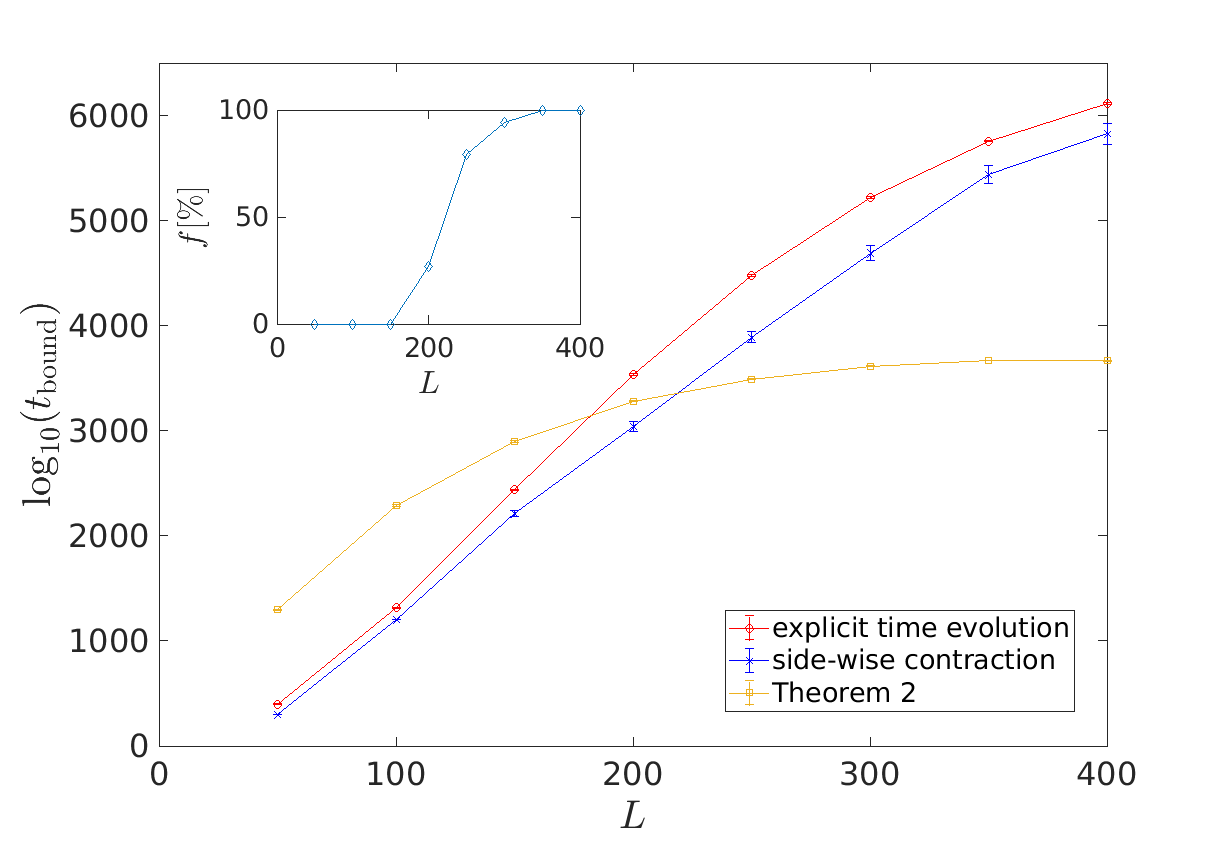}}
\put(-7,175){\textbf{b}}
\end{picture}
	\caption{a: Mean logarithms of the bounds obtained for the classical computation time based on explicit time evolution ($TL^2 2^{2+L^2 - \#(\mathrm{idle \ qubits})}$), side-wise contraction ($2^{1+4FL} L$), Theorem~2, and Algorithm~\ref{algo} averaged over 100 realizations of Sycamore-type quantum circuits. Error bars denote the standard error of the mean of the logarithms. Inset: Same bounds (except the one of Algorithm~\ref{algo}) for large system sizes. %Inset: Fraction $f$ of quantum circuits for which the bound based on Algorithm~\ref{algo} improves over side-wise contraction. This bound improves for almost all realizations over explicit time evolution for $L \geq 6$ (not shown).
		 b: Same for 
	%Mean logarithms of the bounds obtained for the classical computation time based on explicit time evolution ($2^{L^2 - \#(\mathrm{idle \ qubits})}$), side-wise contraction ($2^{4FL} L$), and the second bound of Theorem~2 averaged over 
	100 realizations of random quantum circuits with Poisson-distributed numbers of cavities (save the Algorithm~\ref{algo} results). %Error bars denote the standard error of the mean of the logarithms. 
	Inset: Fraction $f$ of quantum circuits for which the bound of Theorem~2 improves over side-wise contraction.
	It improves over explicit time evolution for all realizations for $L \geq 200$ (not shown).}
% and on average over side-wise contraction for $L \geq 250$. }    
	\label{fig:numerics}
\end{figure}

The practical performance of Algorithm~\ref{algo} can be gathered from Fig. 4 in the main text for short-range instantaneous quantum polynomial-time (IQP) quantum circuits. In Fig.~\ref{fig:numerics}a we show the same data for Sycamore-type quantum circuits~\cite{Bremner2017} acting on an $L \times L$ lattice of qubits. Fig.~\ref{fig:numerics}b displays our numerical data for (2+1)-dimensional random quantum circuits (without the results of Algorithm 1), also for square lattices. 

\bibliography{references}{}

\begin{thebibliography}{40}%
\makeatletter
\providecommand \@ifxundefined [1]{%
 \@ifx{#1\undefined}
}%
\providecommand \@ifnum [1]{%
 \ifnum #1\expandafter \@firstoftwo
 \else \expandafter \@secondoftwo
 \fi
}%
\providecommand \@ifx [1]{%
 \ifx #1\expandafter \@firstoftwo
 \else \expandafter \@secondoftwo
 \fi
}%
\providecommand \natexlab [1]{#1}%
\providecommand \enquote  [1]{``#1''}%
\providecommand \bibnamefont  [1]{#1}%
\providecommand \bibfnamefont [1]{#1}%
\providecommand \citenamefont [1]{#1}%
\providecommand \href@noop [0]{\@secondoftwo}%
\providecommand \href [0]{\begingroup \@sanitize@url \@href}%
\providecommand \@href[1]{\@@startlink{#1}\@@href}%
\providecommand \@@href[1]{\endgroup#1\@@endlink}%
\providecommand \@sanitize@url [0]{\catcode `\\12\catcode `\$12\catcode
  `\&12\catcode `\#12\catcode `\^12\catcode `\_12\catcode `\%12\relax}%
\providecommand \@@startlink[1]{}%
\providecommand \@@endlink[0]{}%
\providecommand \url  [0]{\begingroup\@sanitize@url \@url }%
\providecommand \@url [1]{\endgroup\@href {#1}{\urlprefix }}%
\providecommand \urlprefix  [0]{URL }%
\providecommand \Eprint [0]{\href }%
\providecommand \doibase [0]{https://doi.org/}%
\providecommand \selectlanguage [0]{\@gobble}%
\providecommand \bibinfo  [0]{\@secondoftwo}%
\providecommand \bibfield  [0]{\@secondoftwo}%
\providecommand \translation [1]{[#1]}%
\providecommand \BibitemOpen [0]{}%
\providecommand \bibitemStop [0]{}%
\providecommand \bibitemNoStop [0]{.\EOS\space}%
\providecommand \EOS [0]{\spacefactor3000\relax}%
\providecommand \BibitemShut  [1]{\csname bibitem#1\endcsname}%
\let\auto@bib@innerbib\@empty
%</preamble>
\bibitem [{\citenamefont {Eisert}(2013)}]{Eisert2013}%
  \BibitemOpen
  \bibfield  {author} {\bibinfo {author} {\bibfnamefont {J.}~\bibnamefont
  {Eisert}},\ }\bibfield  {title} {\bibinfo {title} {Entanglement and tensor
  network states,},\ }\href@noop {} {\bibfield  {journal} {\bibinfo  {journal}
  {Model. Simu.}\ }\textbf {\bibinfo {volume} {9}},\ \bibinfo {pages} {520}
  (\bibinfo {year} {2013})}\BibitemShut {NoStop}%
\bibitem [{\citenamefont {Or{\'{u}}s}(2019)}]{Orus2019}%
  \BibitemOpen
  \bibfield  {author} {\bibinfo {author} {\bibfnamefont {R.}~\bibnamefont
  {Or{\'{u}}s}},\ }\bibfield  {title} {\bibinfo {title} {Tensor networks for
  complex quantum systems},\ }\href@noop {} {\bibfield  {journal} {\bibinfo
  {journal} {Nat. Rev. Phys.}\ }\textbf {\bibinfo {volume} {1}},\ \bibinfo
  {pages} {538} (\bibinfo {year} {2019})}\BibitemShut {NoStop}%
\bibitem [{\citenamefont {Tagliacozzo}\ \emph {et~al.}(2014)\citenamefont
  {Tagliacozzo}, \citenamefont {Celi},\ and\ \citenamefont
  {Lewenstein}}]{Tagliacozzo2014}%
  \BibitemOpen
  \bibfield  {author} {\bibinfo {author} {\bibfnamefont {L.}~\bibnamefont
  {Tagliacozzo}}, \bibinfo {author} {\bibfnamefont {A.}~\bibnamefont {Celi}},\
  and\ \bibinfo {author} {\bibfnamefont {M.}~\bibnamefont {Lewenstein}},\
  }\bibfield  {title} {\bibinfo {title} {{Tensor Networks for Lattice Gauge
  Theories with Continuous Groups}},\ }\href@noop {} {\bibfield  {journal}
  {\bibinfo  {journal} {Phys. Rev. X}\ }\textbf {\bibinfo {volume} {4}},\
  \bibinfo {pages} {041024} (\bibinfo {year} {2014})}\BibitemShut {NoStop}%
\bibitem [{\citenamefont {Silvi}\ \emph {et~al.}(2014)\citenamefont {Silvi},
  \citenamefont {Rico}, \citenamefont {Calarco},\ and\ \citenamefont
  {Montangero}}]{Silvi2014}%
  \BibitemOpen
  \bibfield  {author} {\bibinfo {author} {\bibfnamefont {P.}~\bibnamefont
  {Silvi}}, \bibinfo {author} {\bibfnamefont {E.}~\bibnamefont {Rico}},
  \bibinfo {author} {\bibfnamefont {T.}~\bibnamefont {Calarco}},\ and\ \bibinfo
  {author} {\bibfnamefont {S.}~\bibnamefont {Montangero}},\ }\bibfield  {title}
  {\bibinfo {title} {Lattice gauge tensor networks},\ }\href@noop {} {\bibfield
   {journal} {\bibinfo  {journal} {New J. Phys.}\ }\textbf {\bibinfo {volume}
  {16}},\ \bibinfo {pages} {103015} (\bibinfo {year} {2014})}\BibitemShut
  {NoStop}%
\bibitem [{\citenamefont {Zohar}\ \emph {et~al.}(2015)\citenamefont {Zohar},
  \citenamefont {Burrello}, \citenamefont {Wahl},\ and\ \citenamefont
  {Cirac}}]{Zohar2015}%
  \BibitemOpen
  \bibfield  {author} {\bibinfo {author} {\bibfnamefont {E.}~\bibnamefont
  {Zohar}}, \bibinfo {author} {\bibfnamefont {M.}~\bibnamefont {Burrello}},
  \bibinfo {author} {\bibfnamefont {T.~B.}\ \bibnamefont {Wahl}},\ and\
  \bibinfo {author} {\bibfnamefont {J.~I.}\ \bibnamefont {Cirac}},\ }\bibfield
  {title} {\bibinfo {title} {{Fermionic Projected Entangled Pair States and
  Local $U(1)$ Gauge Theories}},\ }\href@noop {} {\bibfield  {journal}
  {\bibinfo  {journal} {Ann. Phys.}\ }\textbf {\bibinfo {volume} {363}},\
  \bibinfo {pages} {385} (\bibinfo {year} {2015})}\BibitemShut {NoStop}%
\bibitem [{\citenamefont {Zohar}\ \emph {et~al.}(2016)\citenamefont {Zohar},
  \citenamefont {Wahl}, \citenamefont {Burrello},\ and\ \citenamefont
  {Cirac}}]{Zohar2016}%
  \BibitemOpen
  \bibfield  {author} {\bibinfo {author} {\bibfnamefont {E.}~\bibnamefont
  {Zohar}}, \bibinfo {author} {\bibfnamefont {T.~B.}\ \bibnamefont {Wahl}},
  \bibinfo {author} {\bibfnamefont {M.}~\bibnamefont {Burrello}},\ and\
  \bibinfo {author} {\bibfnamefont {J.~I.}\ \bibnamefont {Cirac}},\ }\bibfield
  {title} {\bibinfo {title} {{Projected Entangled Pair States with non-Abelian
  gauge symmetries: An $SU(2)$ study}},\ }\href@noop {} {\bibfield  {journal}
  {\bibinfo  {journal} {Ann. Phys.}\ }\textbf {\bibinfo {volume} {374}},\
  \bibinfo {pages} {84} (\bibinfo {year} {2016})}\BibitemShut {NoStop}%
\bibitem [{\citenamefont {Ba{\~n}uls}\ \emph {et~al.}(2019)\citenamefont
  {Ba{\~n}uls}, \citenamefont {Cichy}, \citenamefont {Cirac}, \citenamefont
  {Jansen},\ and\ \citenamefont {K\"{u}hn}}]{Banuls2018}%
  \BibitemOpen
  \bibfield  {author} {\bibinfo {author} {\bibfnamefont {M.~C.}\ \bibnamefont
  {Ba{\~n}uls}}, \bibinfo {author} {\bibfnamefont {K.}~\bibnamefont {Cichy}},
  \bibinfo {author} {\bibfnamefont {J.~I.}\ \bibnamefont {Cirac}}, \bibinfo
  {author} {\bibfnamefont {K.}~\bibnamefont {Jansen}},\ and\ \bibinfo {author}
  {\bibfnamefont {S.}~\bibnamefont {K\"{u}hn}},\ }\bibfield  {title} {\bibinfo
  {title} {{Tensor Networks and their use for Lattice Gauge Theories}},\
  }\bibfield  {booktitle} {\emph {\bibinfo {booktitle}
  {{PoS}({LATTICE}2018)}},\ }\href@noop {} {\ \textbf {\bibinfo {volume} {334}}
  (\bibinfo {year} {2019})}\BibitemShut {NoStop}%
\bibitem [{\citenamefont {Markov}\ and\ \citenamefont
  {Shi}(2008)}]{Markov2008}%
  \BibitemOpen
  \bibfield  {author} {\bibinfo {author} {\bibfnamefont {I.~L.}\ \bibnamefont
  {Markov}}\ and\ \bibinfo {author} {\bibfnamefont {Y.}~\bibnamefont {Shi}},\
  }\bibfield  {title} {\bibinfo {title} {{Simulating Quantum Computation by
  Contracting Tensor Networks}},\ }\href@noop {} {\bibfield  {journal}
  {\bibinfo  {journal} {{SIAM} J. Comp.}\ }\textbf {\bibinfo {volume} {38}},\
  \bibinfo {pages} {963} (\bibinfo {year} {2008})}\BibitemShut {NoStop}%
\bibitem [{\citenamefont {Ferris}\ and\ \citenamefont
  {Poulin}(2014)}]{Ferris2014}%
  \BibitemOpen
  \bibfield  {author} {\bibinfo {author} {\bibfnamefont {A.~J.}\ \bibnamefont
  {Ferris}}\ and\ \bibinfo {author} {\bibfnamefont {D.}~\bibnamefont
  {Poulin}},\ }\bibfield  {title} {\bibinfo {title} {{Tensor Networks and
  Quantum Error Correction}},\ }\href@noop {} {\bibfield  {journal} {\bibinfo
  {journal} {Phys. Rev. Lett.}\ }\textbf {\bibinfo {volume} {113}},\ \bibinfo
  {pages} {030501} (\bibinfo {year} {2014})}\BibitemShut {NoStop}%
\bibitem [{\citenamefont {Guo}\ \emph {et~al.}(2019)\citenamefont {Guo},
  \citenamefont {Liu}, \citenamefont {Xiong}, \citenamefont {Xue},
  \citenamefont {Fu}, \citenamefont {Huang}, \citenamefont {Qiang},
  \citenamefont {Xu}, \citenamefont {Liu}, \citenamefont {Zheng}, \citenamefont
  {Huang}, \citenamefont {Deng}, \citenamefont {Poletti}, \citenamefont {Bao},\
  and\ \citenamefont {Wu}}]{Guo2019}%
  \BibitemOpen
  \bibfield  {author} {\bibinfo {author} {\bibfnamefont {C.}~\bibnamefont
  {Guo}}, \bibinfo {author} {\bibfnamefont {Y.}~\bibnamefont {Liu}}, \bibinfo
  {author} {\bibfnamefont {M.}~\bibnamefont {Xiong}}, \bibinfo {author}
  {\bibfnamefont {S.}~\bibnamefont {Xue}}, \bibinfo {author} {\bibfnamefont
  {X.}~\bibnamefont {Fu}}, \bibinfo {author} {\bibfnamefont {A.}~\bibnamefont
  {Huang}}, \bibinfo {author} {\bibfnamefont {X.}~\bibnamefont {Qiang}},
  \bibinfo {author} {\bibfnamefont {P.}~\bibnamefont {Xu}}, \bibinfo {author}
  {\bibfnamefont {J.}~\bibnamefont {Liu}}, \bibinfo {author} {\bibfnamefont
  {S.}~\bibnamefont {Zheng}}, \bibinfo {author} {\bibfnamefont {H.-L.}\
  \bibnamefont {Huang}}, \bibinfo {author} {\bibfnamefont {M.}~\bibnamefont
  {Deng}}, \bibinfo {author} {\bibfnamefont {D.}~\bibnamefont {Poletti}},
  \bibinfo {author} {\bibfnamefont {W.-S.}\ \bibnamefont {Bao}},\ and\ \bibinfo
  {author} {\bibfnamefont {J.}~\bibnamefont {Wu}},\ }\bibfield  {title}
  {\bibinfo {title} {{General-Purpose Quantum Circuit Simulator with Projected
  Entangled-Pair States and the Quantum Supremacy Frontier}},\ }\href@noop {}
  {\bibfield  {journal} {\bibinfo  {journal} {Phys. Rev. Lett.}\ }\textbf
  {\bibinfo {volume} {123}},\ \bibinfo {pages} {190501} (\bibinfo {year}
  {2019})}\BibitemShut {NoStop}%
\bibitem [{\citenamefont {Pan}\ \emph {et~al.}(2020)\citenamefont {Pan},
  \citenamefont {Zhou}, \citenamefont {Li},\ and\ \citenamefont
  {Zhang}}]{Pan2020}%
  \BibitemOpen
  \bibfield  {author} {\bibinfo {author} {\bibfnamefont {F.}~\bibnamefont
  {Pan}}, \bibinfo {author} {\bibfnamefont {P.}~\bibnamefont {Zhou}}, \bibinfo
  {author} {\bibfnamefont {S.}~\bibnamefont {Li}},\ and\ \bibinfo {author}
  {\bibfnamefont {P.}~\bibnamefont {Zhang}},\ }\bibfield  {title} {\bibinfo
  {title} {{Contracting Arbitrary Tensor Networks: General Approximate
  Algorithm and Applications in Graphical Models and Quantum Circuit
  Simulations}},\ }\href@noop {} {\bibfield  {journal} {\bibinfo  {journal}
  {Phys. Rev. Lett.}\ }\textbf {\bibinfo {volume} {125}},\ \bibinfo {pages}
  {060503} (\bibinfo {year} {2020})}\BibitemShut {NoStop}%
\bibitem [{\citenamefont {Huang}\ \emph {et~al.}(2021)\citenamefont {Huang},
  \citenamefont {Zhang}, \citenamefont {Newman}, \citenamefont {Ni},
  \citenamefont {Ding}, \citenamefont {Cai}, \citenamefont {Gao}, \citenamefont
  {Wang}, \citenamefont {Wu}, \citenamefont {Zhang}, \citenamefont {Ku},
  \citenamefont {Tian}, \citenamefont {Wu}, \citenamefont {Xu}, \citenamefont
  {Yu}, \citenamefont {Yuan}, \citenamefont {Szegedy}, \citenamefont {Shi},
  \citenamefont {Zhao}, \citenamefont {Deng},\ and\ \citenamefont
  {Chen}}]{Huang2021}%
  \BibitemOpen
  \bibfield  {author} {\bibinfo {author} {\bibfnamefont {C.}~\bibnamefont
  {Huang}}, \bibinfo {author} {\bibfnamefont {F.}~\bibnamefont {Zhang}},
  \bibinfo {author} {\bibfnamefont {M.}~\bibnamefont {Newman}}, \bibinfo
  {author} {\bibfnamefont {X.}~\bibnamefont {Ni}}, \bibinfo {author}
  {\bibfnamefont {D.}~\bibnamefont {Ding}}, \bibinfo {author} {\bibfnamefont
  {J.}~\bibnamefont {Cai}}, \bibinfo {author} {\bibfnamefont {X.}~\bibnamefont
  {Gao}}, \bibinfo {author} {\bibfnamefont {T.}~\bibnamefont {Wang}}, \bibinfo
  {author} {\bibfnamefont {F.}~\bibnamefont {Wu}}, \bibinfo {author}
  {\bibfnamefont {G.}~\bibnamefont {Zhang}}, \bibinfo {author} {\bibfnamefont
  {H.-S.}\ \bibnamefont {Ku}}, \bibinfo {author} {\bibfnamefont
  {Z.}~\bibnamefont {Tian}}, \bibinfo {author} {\bibfnamefont {J.}~\bibnamefont
  {Wu}}, \bibinfo {author} {\bibfnamefont {H.}~\bibnamefont {Xu}}, \bibinfo
  {author} {\bibfnamefont {H.}~\bibnamefont {Yu}}, \bibinfo {author}
  {\bibfnamefont {B.}~\bibnamefont {Yuan}}, \bibinfo {author} {\bibfnamefont
  {M.}~\bibnamefont {Szegedy}}, \bibinfo {author} {\bibfnamefont
  {Y.}~\bibnamefont {Shi}}, \bibinfo {author} {\bibfnamefont {H.-H.}\
  \bibnamefont {Zhao}}, \bibinfo {author} {\bibfnamefont {C.}~\bibnamefont
  {Deng}},\ and\ \bibinfo {author} {\bibfnamefont {J.}~\bibnamefont {Chen}},\
  }\bibfield  {title} {\bibinfo {title} {Efficient parallelization of tensor
  network contraction for simulating quantum computation},\ }\href@noop {}
  {\bibfield  {journal} {\bibinfo  {journal} {Nat. Comp. Sci.}\ }\textbf
  {\bibinfo {volume} {1}},\ \bibinfo {pages} {578} (\bibinfo {year}
  {2021})}\BibitemShut {NoStop}%
\bibitem [{\citenamefont {Farrelly}\ \emph {et~al.}(2021)\citenamefont
  {Farrelly}, \citenamefont {Harris}, \citenamefont {McMahon},\ and\
  \citenamefont {Stace}}]{Farrelly2021}%
  \BibitemOpen
  \bibfield  {author} {\bibinfo {author} {\bibfnamefont {T.}~\bibnamefont
  {Farrelly}}, \bibinfo {author} {\bibfnamefont {R.~J.}\ \bibnamefont
  {Harris}}, \bibinfo {author} {\bibfnamefont {N.~A.}\ \bibnamefont
  {McMahon}},\ and\ \bibinfo {author} {\bibfnamefont {T.~M.}\ \bibnamefont
  {Stace}},\ }\bibfield  {title} {\bibinfo {title} {{Tensor-Network Codes}},\
  }\href@noop {} {\bibfield  {journal} {\bibinfo  {journal} {Phys. Rev. Lett.}\
  }\textbf {\bibinfo {volume} {127}},\ \bibinfo {pages} {040507} (\bibinfo
  {year} {2021})}\BibitemShut {NoStop}%
\bibitem [{\citenamefont {Levental}()}]{Levental2021}%
  \BibitemOpen
  \bibfield  {author} {\bibinfo {author} {\bibfnamefont {M.}~\bibnamefont
  {Levental}},\ }\bibfield  {title} {\bibinfo {title} {{Tensor Networks for
  Simulating Quantum Circuits on FPGAs}},\ }\href@noop {} {\bibinfo  {journal}
  {arXiv:2108.06831}\ }\BibitemShut {NoStop}%
\bibitem [{\citenamefont {Napp}\ \emph {et~al.}(2022)\citenamefont {Napp},
  \citenamefont {{La Placa}}, \citenamefont {Dalzell}, \citenamefont
  {Brand{\~{a}}o},\ and\ \citenamefont {Harrow}}]{Napp2022}%
  \BibitemOpen
\bibfield  {journal} {  }\bibfield  {author} {\bibinfo {author} {\bibfnamefont
  {J.~C.}\ \bibnamefont {Napp}}, \bibinfo {author} {\bibfnamefont {R.~L.}\
  \bibnamefont {{La Placa}}}, \bibinfo {author} {\bibfnamefont {A.~M.}\
  \bibnamefont {Dalzell}}, \bibinfo {author} {\bibfnamefont {F.~G. S.~L.}\
  \bibnamefont {Brand{\~{a}}o}},\ and\ \bibinfo {author} {\bibfnamefont
  {A.~W.}\ \bibnamefont {Harrow}},\ }\bibfield  {title} {\bibinfo {title}
  {{Efficient Classical Simulation of Random Shallow 2D Quantum Circuits}},\
  }\href@noop {} {\bibfield  {journal} {\bibinfo  {journal} {Phys. Rev. X}\
  }\textbf {\bibinfo {volume} {12}},\ \bibinfo {pages} {021021} (\bibinfo
  {year} {2022})}\BibitemShut {NoStop}%
\bibitem [{\citenamefont {Stoudenmire}\ and\ \citenamefont
  {Schwab}(2016)}]{Stoudenmire2016}%
  \BibitemOpen
  \bibfield  {author} {\bibinfo {author} {\bibfnamefont {E.}~\bibnamefont
  {Stoudenmire}}\ and\ \bibinfo {author} {\bibfnamefont {D.~J.}\ \bibnamefont
  {Schwab}},\ }\bibfield  {title} {\bibinfo {title} {{Supervised Learning with
  Tensor Networks}},\ }\href@noop {} {\bibfield  {journal} {\bibinfo  {journal}
  {Adv. Neur. Info. Proc. Sys.}\ }\textbf {\bibinfo {volume} {29}},\ \bibinfo
  {pages} {4799} (\bibinfo {year} {2016})}\BibitemShut {NoStop}%
\bibitem [{\citenamefont {Arute}\ \emph {et~al.}(2019)\citenamefont {Arute},
  \citenamefont {Arya}, \citenamefont {Babbush}, \citenamefont {Bacon},
  \citenamefont {Bardin} \emph {et~al.}}]{Arute2019googlesycamore}%
  \BibitemOpen
  \bibfield  {author} {\bibinfo {author} {\bibfnamefont {F.}~\bibnamefont
  {Arute}}, \bibinfo {author} {\bibfnamefont {K.}~\bibnamefont {Arya}},
  \bibinfo {author} {\bibfnamefont {R.}~\bibnamefont {Babbush}}, \bibinfo
  {author} {\bibfnamefont {D.}~\bibnamefont {Bacon}}, \bibinfo {author}
  {\bibfnamefont {J.~C.}\ \bibnamefont {Bardin}}, \emph {et~al.},\ }\bibfield
  {title} {\bibinfo {title} {{Quantum supremacy using a programmable
  superconducting processor}},\ }\href@noop {} {\bibfield  {journal} {\bibinfo
  {journal} {\nat}\ }\textbf {\bibinfo {volume} {574}},\ \bibinfo {pages} {505}
  (\bibinfo {year} {2019})}\BibitemShut {NoStop}%
\bibitem [{\citenamefont {Guo}\ \emph {et~al.}(2021)\citenamefont {Guo},
  \citenamefont {Zhao},\ and\ \citenamefont {Huang}}]{Guo2021}%
  \BibitemOpen
  \bibfield  {author} {\bibinfo {author} {\bibfnamefont {C.}~\bibnamefont
  {Guo}}, \bibinfo {author} {\bibfnamefont {Y.}~\bibnamefont {Zhao}},\ and\
  \bibinfo {author} {\bibfnamefont {H.-L.}\ \bibnamefont {Huang}},\ }\bibfield
  {title} {\bibinfo {title} {{Verifying Random Quantum Circuits with Arbitrary
  Geometry Using Tensor Network States Algorithm}},\ }\href@noop {} {\bibfield
  {journal} {\bibinfo  {journal} {Phys. Rev. Lett.}\ }\textbf {\bibinfo
  {volume} {126}},\ \bibinfo {pages} {070502} (\bibinfo {year}
  {2021})}\BibitemShut {NoStop}%
\bibitem [{\citenamefont {Gray}\ and\ \citenamefont
  {Kourtis}(2021{\natexlab{a}})}]{Gray2021}%
  \BibitemOpen
  \bibfield  {author} {\bibinfo {author} {\bibfnamefont {J.}~\bibnamefont
  {Gray}}\ and\ \bibinfo {author} {\bibfnamefont {S.}~\bibnamefont {Kourtis}},\
  }\bibfield  {title} {\bibinfo {title} {Hyper-optimized tensor network
  contraction},\ }\href@noop {} {\bibfield  {journal} {\bibinfo  {journal}
  {Quantum}\ }\textbf {\bibinfo {volume} {5}},\ \bibinfo {pages} {410}
  (\bibinfo {year} {2021}{\natexlab{a}})}\BibitemShut {NoStop}%
\bibitem [{\citenamefont {Huang}\ \emph {et~al.}()\citenamefont {Huang},
  \citenamefont {Zhang}, \citenamefont {Newman}, \citenamefont {Cai},
  \citenamefont {Gao}, \citenamefont {Tian}, \citenamefont {Wu}, \citenamefont
  {Xu}, \citenamefont {Yu}, \citenamefont {Yuan}, \citenamefont {Szegedy},
  \citenamefont {Shi},\ and\ \citenamefont {Chen}}]{Alibaba}%
  \BibitemOpen
  \bibfield  {author} {\bibinfo {author} {\bibfnamefont {C.}~\bibnamefont
  {Huang}}, \bibinfo {author} {\bibfnamefont {F.}~\bibnamefont {Zhang}},
  \bibinfo {author} {\bibfnamefont {M.}~\bibnamefont {Newman}}, \bibinfo
  {author} {\bibfnamefont {J.}~\bibnamefont {Cai}}, \bibinfo {author}
  {\bibfnamefont {X.}~\bibnamefont {Gao}}, \bibinfo {author} {\bibfnamefont
  {Z.}~\bibnamefont {Tian}}, \bibinfo {author} {\bibfnamefont {J.}~\bibnamefont
  {Wu}}, \bibinfo {author} {\bibfnamefont {H.}~\bibnamefont {Xu}}, \bibinfo
  {author} {\bibfnamefont {H.}~\bibnamefont {Yu}}, \bibinfo {author}
  {\bibfnamefont {B.}~\bibnamefont {Yuan}}, \bibinfo {author} {\bibfnamefont
  {M.}~\bibnamefont {Szegedy}}, \bibinfo {author} {\bibfnamefont
  {Y.}~\bibnamefont {Shi}},\ and\ \bibinfo {author} {\bibfnamefont
  {J.}~\bibnamefont {Chen}},\ }\bibfield  {title} {\bibinfo {title} {{Classical
  Simulation of Quantum Supremacy Circuits}},\ }\href@noop {} {\bibinfo
  {journal} {arXiv:2005.06787}\ }\BibitemShut {NoStop}%
\bibitem [{\citenamefont {Pan}\ \emph {et~al.}(2022)\citenamefont {Pan},
  \citenamefont {Chen},\ and\ \citenamefont {Zhang}}]{pan2021solving}%
  \BibitemOpen
\bibfield  {journal} {  }\bibfield  {author} {\bibinfo {author} {\bibfnamefont
  {F.}~\bibnamefont {Pan}}, \bibinfo {author} {\bibfnamefont {K.}~\bibnamefont
  {Chen}},\ and\ \bibinfo {author} {\bibfnamefont {P.}~\bibnamefont {Zhang}},\
  }\bibfield  {title} {\bibinfo {title} {{Solving the sampling problem of the
  Sycamore quantum supremacy circuits}},\ }\href@noop {} {\bibfield  {journal}
  {\bibinfo  {journal} {Phys. Rev. Lett.}\ }\textbf {\bibinfo {volume} {129}},\
  \bibinfo {pages} {090502} (\bibinfo {year} {2022})}\BibitemShut {NoStop}%
\bibitem [{\citenamefont {Pan}\ and\ \citenamefont
  {Zhang}()}]{pan2021simulating}%
  \BibitemOpen
  \bibfield  {author} {\bibinfo {author} {\bibfnamefont {F.}~\bibnamefont
  {Pan}}\ and\ \bibinfo {author} {\bibfnamefont {P.}~\bibnamefont {Zhang}},\
  }\bibfield  {title} {\bibinfo {title} {{Simulating the Sycamore quantum
  supremacy circuits}},\ }\href@noop {} {\bibinfo  {journal}
  {arXiv:2103.03074}\ }\BibitemShut {NoStop}%
\bibitem [{\citenamefont {Liu}\ \emph {et~al.}(2021)\citenamefont {Liu},
  \citenamefont {Liu}, \citenamefont {Li}, \citenamefont {Fu}, \citenamefont
  {Yang}, \citenamefont {Song}, \citenamefont {Zhao}, \citenamefont {Wang},
  \citenamefont {Peng}, \citenamefont {Chen} \emph {et~al.}}]{liu2021closing}%
  \BibitemOpen
\bibfield  {journal} {  }\bibfield  {author} {\bibinfo {author} {\bibfnamefont
  {Y.}~\bibnamefont {Liu}}, \bibinfo {author} {\bibfnamefont {X.}~\bibnamefont
  {Liu}}, \bibinfo {author} {\bibfnamefont {F.}~\bibnamefont {Li}}, \bibinfo
  {author} {\bibfnamefont {H.}~\bibnamefont {Fu}}, \bibinfo {author}
  {\bibfnamefont {Y.}~\bibnamefont {Yang}}, \bibinfo {author} {\bibfnamefont
  {J.}~\bibnamefont {Song}}, \bibinfo {author} {\bibfnamefont {P.}~\bibnamefont
  {Zhao}}, \bibinfo {author} {\bibfnamefont {Z.}~\bibnamefont {Wang}}, \bibinfo
  {author} {\bibfnamefont {D.}~\bibnamefont {Peng}}, \bibinfo {author}
  {\bibfnamefont {H.}~\bibnamefont {Chen}}, \emph {et~al.},\ }\bibfield
  {title} {\bibinfo {title} {Closing the ``quantum supremacy'' gap: achieving
  real-time simulation of a random quantum circuit using a new sunway
  supercomputer},\ }in\ \href@noop {} {\emph {\bibinfo {booktitle} {Proceedings
  of the International Conference for High Performance Computing, Networking,
  Storage and Analysis}}}\ (\bibinfo {year} {2021})\ pp.\ \bibinfo {pages}
  {1--12}\BibitemShut {NoStop}%
\bibitem [{\citenamefont {Dumitrescu}\ \emph {et~al.}(2018)\citenamefont
  {Dumitrescu}, \citenamefont {Fisher}, \citenamefont {Goodrich}, \citenamefont
  {Humble},\ and\ \citenamefont {Wright}}]{treewidth1}%
  \BibitemOpen
  \bibfield  {author} {\bibinfo {author} {\bibfnamefont {E.~F.}\ \bibnamefont
  {Dumitrescu}}, \bibinfo {author} {\bibfnamefont {A.~L.}\ \bibnamefont
  {Fisher}}, \bibinfo {author} {\bibfnamefont {T.~D.}\ \bibnamefont
  {Goodrich}}, \bibinfo {author} {\bibfnamefont {B.~D.}\ \bibnamefont {Humble},
  \bibfnamefont {Travis S.~Sullivan}},\ and\ \bibinfo {author} {\bibfnamefont
  {A.~L.}\ \bibnamefont {Wright}},\ }\bibfield  {title} {\bibinfo {title}
  {Benchmarking treewidth as a practical component of tensor network
  simulations},\ }\href@noop {} {\bibfield  {journal} {\bibinfo  {journal}
  {Plos One}\ }\textbf {\bibinfo {volume} {13}},\ \bibinfo {pages} {12}
  (\bibinfo {year} {2018})}\BibitemShut {NoStop}%
\bibitem [{\citenamefont {Dudek}\ \emph {et~al.}()\citenamefont {Dudek},
  \citenamefont {Due\~{n}as Osorio},\ and\ \citenamefont {Vardi}}]{treewidth2}%
  \BibitemOpen
  \bibfield  {author} {\bibinfo {author} {\bibfnamefont {J.~M.}\ \bibnamefont
  {Dudek}}, \bibinfo {author} {\bibfnamefont {L.}~\bibnamefont {Due\~{n}as
  Osorio}},\ and\ \bibinfo {author} {\bibfnamefont {M.~Y.}\ \bibnamefont
  {Vardi}},\ }\bibfield  {title} {\bibinfo {title} {{Efficient Contraction of
  Large Tensor Networks for Weighted Model Counting through Graph
  Decompositions}},\ }\href@noop {} {\bibinfo  {journal} {arXiv:1908.04381}\
  }\BibitemShut {NoStop}%
\bibitem [{\citenamefont {Ayral}\ \emph {et~al.}()\citenamefont {Ayral},
  \citenamefont {Louvet}, \citenamefont {Zhou}, \citenamefont {Lambert},
  \citenamefont {Stoudenmire},\ and\ \citenamefont
  {Waintal}}]{ayral2022density}%
  \BibitemOpen
\bibfield  {journal} {  }\bibfield  {author} {\bibinfo {author} {\bibfnamefont
  {T.}~\bibnamefont {Ayral}}, \bibinfo {author} {\bibfnamefont
  {T.}~\bibnamefont {Louvet}}, \bibinfo {author} {\bibfnamefont
  {Y.}~\bibnamefont {Zhou}}, \bibinfo {author} {\bibfnamefont {C.}~\bibnamefont
  {Lambert}}, \bibinfo {author} {\bibfnamefont {E.~M.}\ \bibnamefont
  {Stoudenmire}},\ and\ \bibinfo {author} {\bibfnamefont {X.}~\bibnamefont
  {Waintal}},\ }\bibfield  {title} {\bibinfo {title} {A density-matrix
  renormalisation group algorithm for simulating quantum circuits with a finite
  fidelity},\ }\href@noop {} {\bibinfo  {journal} {arXiv:2207.05612}\
  }\BibitemShut {NoStop}%
\bibitem [{\citenamefont {Lipton}\ and\ \citenamefont
  {Tarjan}(1979)}]{Lipton1979planar}%
  \BibitemOpen
\bibfield  {journal} {  }\bibfield  {author} {\bibinfo {author} {\bibfnamefont
  {R.~J.}\ \bibnamefont {Lipton}}\ and\ \bibinfo {author} {\bibfnamefont
  {R.~E.}\ \bibnamefont {Tarjan}},\ }\bibfield  {title} {\bibinfo {title} {{A
  Separator Theorem for Planar Graphs}},\ }\href@noop {} {\bibfield  {journal}
  {\bibinfo  {journal} {{SIAM} Journal on Applied Mathematics}\ }\textbf
  {\bibinfo {volume} {36}},\ \bibinfo {pages} {177} (\bibinfo {year}
  {1979})}\BibitemShut {NoStop}%
\bibitem [{\citenamefont {Miller}\ \emph {et~al.}(1997)\citenamefont {Miller},
  \citenamefont {Teng}, \citenamefont {Thurston},\ and\ \citenamefont
  {Vavasis}}]{miller1997separators}%
  \BibitemOpen
  \bibfield  {author} {\bibinfo {author} {\bibfnamefont {G.~L.}\ \bibnamefont
  {Miller}}, \bibinfo {author} {\bibfnamefont {S.-H.}\ \bibnamefont {Teng}},
  \bibinfo {author} {\bibfnamefont {W.}~\bibnamefont {Thurston}},\ and\
  \bibinfo {author} {\bibfnamefont {S.~A.}\ \bibnamefont {Vavasis}},\
  }\bibfield  {title} {\bibinfo {title} {Separators for sphere-packings and
  nearest neighbor graphs},\ }\href@noop {} {\bibfield  {journal} {\bibinfo
  {journal} {Journal of the ACM (JACM)}\ }\textbf {\bibinfo {volume} {44}},\
  \bibinfo {pages} {1} (\bibinfo {year} {1997})}\BibitemShut {NoStop}%
\bibitem [{\citenamefont {Kourtis}\ \emph {et~al.}(2019)\citenamefont
  {Kourtis}, \citenamefont {Chamon}, \citenamefont {Mucciolo},\ and\
  \citenamefont {Ruckenstein}}]{Kourtis2019couting}%
  \BibitemOpen
  \bibfield  {author} {\bibinfo {author} {\bibfnamefont {S.}~\bibnamefont
  {Kourtis}}, \bibinfo {author} {\bibfnamefont {C.}~\bibnamefont {Chamon}},
  \bibinfo {author} {\bibfnamefont {E.}~\bibnamefont {Mucciolo}},\ and\
  \bibinfo {author} {\bibfnamefont {A.}~\bibnamefont {Ruckenstein}},\
  }\bibfield  {title} {\bibinfo {title} {Fast counting with tensor networks},\
  }\href@noop {} {\bibfield  {journal} {\bibinfo  {journal} {{SciPost}
  Physics}\ }\textbf {\bibinfo {volume} {7}} (\bibinfo {year}
  {2019})}\BibitemShut {NoStop}%
\bibitem [{\citenamefont {Liu}()}]{Liu2020}%
  \BibitemOpen
  \bibfield  {author} {\bibinfo {author} {\bibfnamefont {Y.}~\bibnamefont
  {Liu}},\ }\bibfield  {title} {\bibinfo {title} {{The Complexity of
  Contracting Planar Tensor Network}},\ }\href@noop {} {\bibinfo  {journal}
  {arXiv:2001.10204}\ }\BibitemShut {NoStop}%
\bibitem [{\citenamefont {Huang}\ \emph {et~al.}(2020)\citenamefont {Huang},
  \citenamefont {Newman},\ and\ \citenamefont {Szegedy}}]{huang2020explicit}%
  \BibitemOpen
\bibfield  {journal} {  }\bibfield  {author} {\bibinfo {author} {\bibfnamefont
  {C.}~\bibnamefont {Huang}}, \bibinfo {author} {\bibfnamefont
  {M.}~\bibnamefont {Newman}},\ and\ \bibinfo {author} {\bibfnamefont
  {M.}~\bibnamefont {Szegedy}},\ }\bibfield  {title} {\bibinfo {title}
  {Explicit lower bounds on strong quantum simulation},\ }\href@noop {}
  {\bibfield  {journal} {\bibinfo  {journal} {IEEE Transactions on Information
  Theory}\ }\textbf {\bibinfo {volume} {66}},\ \bibinfo {pages} {5585}
  (\bibinfo {year} {2020})}\BibitemShut {NoStop}%
\bibitem [{\citenamefont {Gray}\ and\ \citenamefont
  {Kourtis}(2021{\natexlab{b}})}]{gray2020hyper}%
  \BibitemOpen
  \bibfield  {author} {\bibinfo {author} {\bibfnamefont {J.}~\bibnamefont
  {Gray}}\ and\ \bibinfo {author} {\bibfnamefont {S.}~\bibnamefont {Kourtis}},\
  }\bibfield  {title} {\bibinfo {title} {Hyper-optimized tensor network
  contraction},\ }\href {https://doi.org/10.22331/q-2021-03-15-410} {\bibfield
  {journal} {\bibinfo  {journal} {Quantum}\ }\textbf {\bibinfo {volume} {5}},\
  \bibinfo {pages} {410} (\bibinfo {year} {2021}{\natexlab{b}})}\BibitemShut
  {NoStop}%
\bibitem [{\citenamefont {Smith}\ and\ \citenamefont
  {Wormald}(1998)}]{Smith1998cubesep}%
  \BibitemOpen
  \bibfield  {author} {\bibinfo {author} {\bibfnamefont {W.}~\bibnamefont
  {Smith}}\ and\ \bibinfo {author} {\bibfnamefont {N.}~\bibnamefont
  {Wormald}},\ }\bibfield  {title} {\bibinfo {title} {Geometric separator
  theorems and applications},\ }in\ \href@noop {} {\emph {\bibinfo {booktitle}
  {Proceedings 39th Annual Symposium on Foundations of Computer Science (Cat.
  No.98CB36280)}}}\ (\bibinfo  {publisher} {{IEEE} Comput. Soc},\ \bibinfo
  {year} {1998})\BibitemShut {NoStop}%
\bibitem [{\citenamefont {Djidjev}\ and\ \citenamefont
  {Venkatesan}(1997)}]{Djidjev1997improved}%
  \BibitemOpen
  \bibfield  {author} {\bibinfo {author} {\bibfnamefont {H.~N.}\ \bibnamefont
  {Djidjev}}\ and\ \bibinfo {author} {\bibfnamefont {S.~M.}\ \bibnamefont
  {Venkatesan}},\ }\bibfield  {title} {\bibinfo {title} {Reduced constants for
  simple cycle graph separation},\ }\href@noop {} {\bibfield  {journal}
  {\bibinfo  {journal} {Acta Informatica}\ }\textbf {\bibinfo {volume} {34}},\
  \bibinfo {pages} {231} (\bibinfo {year} {1997})}\BibitemShut {NoStop}%
\bibitem [{\citenamefont {Bremner}\ \emph {et~al.}(2017)\citenamefont
  {Bremner}, \citenamefont {Montanaro},\ and\ \citenamefont
  {Shepherd}}]{Bremner2017}%
  \BibitemOpen
  \bibfield  {author} {\bibinfo {author} {\bibfnamefont {M.~J.}\ \bibnamefont
  {Bremner}}, \bibinfo {author} {\bibfnamefont {A.}~\bibnamefont {Montanaro}},\
  and\ \bibinfo {author} {\bibfnamefont {D.~J.}\ \bibnamefont {Shepherd}},\
  }\bibfield  {title} {\bibinfo {title} {Achieving quantum supremacy with
  sparse and noisy commuting quantum computations},\ }\href@noop {} {\bibfield
  {journal} {\bibinfo  {journal} {Quantum}\ }\textbf {\bibinfo {volume} {1}},\
  \bibinfo {pages} {8} (\bibinfo {year} {2017})}\BibitemShut {NoStop}%
\bibitem [{\citenamefont {Kesten}(1980)}]{Kesten1980}%
  \BibitemOpen
  \bibfield  {author} {\bibinfo {author} {\bibfnamefont {H.}~\bibnamefont
  {Kesten}},\ }\bibfield  {title} {\bibinfo {title} {The critical probability
  of bond percolation on the square lattice equals 1/2},\ }\href
  {https://doi.org/10.1007/bf01197577} {\bibfield  {journal} {\bibinfo
  {journal} {Commun. Math. Phys.}\ }\textbf {\bibinfo {volume} {74}},\ \bibinfo
  {pages} {41} (\bibinfo {year} {1980})}\BibitemShut {NoStop}%
\bibitem [{\citenamefont {Edelsbrunner}(1987)}]{Edelsbrunner1987}%
  \BibitemOpen
  \bibfield  {author} {\bibinfo {author} {\bibfnamefont {H.}~\bibnamefont
  {Edelsbrunner}},\ }\href@noop {} {\emph {\bibinfo {title} {{Algorithms in
  Combinatorial Geometry}}}}\ (\bibinfo  {publisher} {Springer Berlin
  Heidelberg},\ \bibinfo {year} {1987})\BibitemShut {NoStop}%
\bibitem [{\citenamefont {Vapnik}\ and\ \citenamefont
  {Chervonenkis}(1971)}]{Vapnik1971}%
  \BibitemOpen
  \bibfield  {author} {\bibinfo {author} {\bibfnamefont {V.~N.}\ \bibnamefont
  {Vapnik}}\ and\ \bibinfo {author} {\bibfnamefont {A.~Y.}\ \bibnamefont
  {Chervonenkis}},\ }\href@noop {} {\bibfield  {journal} {\bibinfo  {journal}
  {Theory Prob. Appl.}\ }\textbf {\bibinfo {volume} {16}},\ \bibinfo {pages}
  {264} (\bibinfo {year} {1971})}\BibitemShut {NoStop}%
\bibitem [{\citenamefont {Haussler}\ and\ \citenamefont
  {Welzl}(1987)}]{Haussler1987}%
  \BibitemOpen
  \bibfield  {author} {\bibinfo {author} {\bibfnamefont {D.}~\bibnamefont
  {Haussler}}\ and\ \bibinfo {author} {\bibfnamefont {E.}~\bibnamefont
  {Welzl}},\ }\bibfield  {title} {\bibinfo {title} {{$\varepsilon$}-nets and
  simplex range queries},\ }\href@noop {} {\bibfield  {journal} {\bibinfo
  {journal} {Discrete {\&} Computational Geometry}\ }\textbf {\bibinfo {volume}
  {2}},\ \bibinfo {pages} {127} (\bibinfo {year} {1987})}\BibitemShut {NoStop}%
\bibitem [{\citenamefont {Teng}(1991)}]{Teng1991}%
  \BibitemOpen
  \bibfield  {author} {\bibinfo {author} {\bibfnamefont {S.-H.}\ \bibnamefont
  {Teng}},\ }\emph {\bibinfo {title} {{Point, Spheres, and Separators: A
  unified geometric approach to graph partitioning}}},\ \href@noop {} {Ph.D.
  thesis},\ \bibinfo  {school} {CMU-CS-91-184, School of Computer Science},
  \bibinfo {address} {Carnegie-Mellon Univ., Pittsburgh, Pa.} (\bibinfo {year}
  {1991})\BibitemShut {NoStop}%
\end{thebibliography}%

\end{document}